\documentclass[usenatbib, graybox, nosecnum]{svmult}

\usepackage{mathptmx}       
\usepackage{helvet}         
\usepackage{courier}        
\usepackage{type1cm}        
\usepackage{makeidx}         
\usepackage{graphicx}        
\usepackage{multicol}        
\usepackage[bottom]{footmisc}
\usepackage{hyperref}        
\usepackage{soul}            
\hypersetup{colorlinks=true,urlcolor=blue}
\usepackage{color}
\usepackage{xcolor}

\usepackage{enumitem}
\usepackage{amsmath}
\usepackage{amssymb}
\usepackage{geometry}
\usepackage{pdflscape}

\usepackage[square,numbers]{natbib}
\makeindex             

\newcommand{\unitspace}{\,}
\newcommand{\km}{\ensuremath{\unitspace \mathrm{km}}}
\newcommand{\cm}{\ensuremath{\unitspace \mathrm{cm}}}
\newcommand{\gram}{\ensuremath{\unitspace \mathrm{g}}}
\newcommand{\Hz}{\ensuremath{\unitspace \mathrm{Hz}}}
\newcommand{\second}{\ensuremath{\unitspace \mathrm{s}}}
\newcommand{\erg}{\ensuremath{\unitspace \mathrm{erg}}}
\newcommand{\kev}{\ensuremath{\unitspace \mathrm{keV}}}
\newcommand{\eV}{\ensuremath{\unitspace \mathrm{eV}}}
\newcommand{\ergs}{\ensuremath{\unitspace \mathrm{erg} \unitspace \second^{-1}}}

\newcommand{\yr}{\ensuremath{\unitspace \mathrm{yr}}}
\newcommand{\gauss}{\ensuremath{\unitspace \mathrm{G}}}
\newcommand{\Kelvin}{\ensuremath{\unitspace \mathrm{K}}}
\newcommand{\Msun}{\ensuremath{\unitspace \mathrm{M}_{\odot}}}

\newcommand{\rb}{\ensuremath{\bar{r}}}
\newcommand{\ub}{\ensuremath{\bar{u}}}

\newcommand{\ud}{\ensuremath{\mathrm{d}}}

\renewcommand{\vec}[1]{\mathbf{#1}}
\newcommand{\nvec}[1]{\hat{\vec{#1}}}
\newcommand{\Ten}[2]{\ensuremath{#1 \times 10^{#2}} }

\newcommand{\apj}{ApJ}
\newcommand{\apjl}{ApJ}
\newcommand{\apjs}{ApJS}
\newcommand{\prl}{Phys.~Rev.~Lett.}
\newcommand{\nat}{Nature}
\newcommand{\ssr}{Space~Sci.~Rev.}
\newcommand{\mnras}{MNRAS}
\newcommand{\araa}{ARA\&A}
\newcommand{\aap}{A\&A}
\newcommand{\pre}{Phys.~Rev.~E}
\newcommand{\aapr}{A\&A~Rev.}
\newcommand{\prd}{Phys.~Rev.~D}
\newcommand{\aplett}{Astrophys.~Lett.}
\newcommand{\apss}{Ap\&SS}

\begin{document}
\def\hbar{{\mathchar'26\mkern-9mu h}}

\title*{Fundamental physics with neutron stars}

\author{Joonas N\"attil\"a\thanks{corresponding author} and Jari J.E. Kajava}

\institute{Joonas N\"attil\"a \at Center for Computational Astrophysics, Flatiron Institute, 162 Fifth Avenue, New York 10010, USA, \\
Physics Department \& Columbia Astrophysics Laboratory, Columbia University, 538 West 120th Street, New York 10027, USA.\\ 
\email{jnattila@flatironinstitute.org}
\and Jari J. E. Kajava \at Serco for ESA, ESA/ESAC, 28692 Villanueva de la Ca\~{n}ada, Madrid, Spain, \\
Department of Physics and Astronomy, 20014 University of Turku, Finland.\\ \email{jari.kajava@ext.esa.int}}


\maketitle
\abstract{
Neutron stars are rich laboratories of multiple branches of modern physics.
These include gravitational physics, nuclear and particle physics, (quantum) electrodynamics, and plasma astrophysics.
In this chapter, we present the pioneering theoretical studies and the pivotal historical observations on which our understanding of neutron stars is based on.
Then, we discuss the usage of neutron stars as probes of fundamental theories of physics.
}
\section{Keywords} 
Neutron stars ---
Pulsars ---
Magnetars ---
Mergers ---
Dense matter ---
Gravitational Physics ---
Particle Physics ---
Nuclear Physics ---
Electrodynamics ----
Plasma Physics


\section{Introduction}

Neutron stars are extremely dense, rapidly rotating, superfluid magnets. 
They shine to us across a broad electromagnetic spectrum from low-frequency radio bands to high-energy X-rays and gamma-rays.
Some of them produce continuous radiation, some show pulsations, and some explode in flares.
This activity is powered by their huge gravitational, magnetic, rotational, and chemical energy.
These enormous reservoirs of energy, in turn, drive exotic phenomena encompassing many fields of modern physics, from the quantum realm to space-time-deforming general relativity.
Most of the observed phenomena from neutron stars are so extreme that they can not be studied in contemporary terrestrial laboratories.
This makes neutron stars \textit{the cosmic laboratories of extreme physics}.

A canonical neutron star has a radius of $R \approx 12\km$ and a mass of $M \approx 1.5 \Msun$; 
the observed masses range from about $1.1$ to $2.1\Msun$, and theoretically supported radii range from about $10$ to $15\km$.
This makes them extremely dense astrophysical objects capable of significantly distorting the local space-time---hence, general relativity is needed when describing the physical laws around them.
Neutron stars are also observed to rotate with spin periods ranging from $P \sim 0.001$ to $10 \second$.
When such a tightly compressed object is rapidly rotating, it stores massive amounts of angular momentum that will result in additional ``twisting'' of the space-time ``fabric'', in addition to the more usual curving effect caused by the concentrated mass. 
The first section of this chapter describes the gravitational physics of neutron stars.

The large density means that the matter inside neutron stars is under tremendous pressure. 
The mean density inside the star, $\bar{\rho} \sim 10^{15} \gram \cm^{-3}$,  is comparable to the density inside the atomic nuclei.
The matter is compressed into such a small volume that quantum mechanical effects define its behavior.
How the matter responds to this squeezing, as the particles try to counter the compressing force, is encapsulated in the---still unknown---equation of state (EOS) of the cold, dense matter. 
The nuclear and particle physics of neutron stars are described in the second section of this chapter.

Neutron stars have strong magnetic fields, ranging from $B \sim 10^8$ to $10^{15} \gauss$.
As the star spins, the topology of this dynamic magnetosphere changes.
The physics of the magnetosphere is governed by both electromagnetism, describing the dynamical evolution of the electric and magnetic fields, and quantum electrodynamics (QED), describing the coupling of the electromagnetic fields and particles.
The electrodynamics, QED phenomena, and plasma physics of neutron stars are discussed in the third and fourth sections of this chapter.


\subsection{Formation of neutron stars}

The idea of stars composed of neutrons originates from Lev Landau \citep{landau_1932, yakovlev_2013};
he was the first to speculate on the possibility of squeezing the matter inside regular stars so tightly that quantum effects would start to dominate:
\textit{"...the density of the matter becomes so great that atomic nuclei come in close contact, forming one gigantic nucleus"}.
Interestingly, this idea was coined even before the existence of neutrons was known.%
\footnote{%
Existence of electrically neutral particles with a mass $m_n \approx m_p \approx 5 \times 10^{-25} \gram$ was proven by Chadwick in 1932 \citep{chadwick_1932a, chadwick_1932b}.
}
Similar ideas were later proposed also by Walter Baade and Fritz Zwicky \citep{baade_1934}---they were even bold enough to suggest that neutron stars might form in supernovae.
This turned out to be exactly right!

Neutron stars are now known to be born from a collapse of a main-sequence star, slightly more massive than our own Sun, with a mass $M_\star \sim 10 \Msun$, where $\Msun \approx \Ten{1.99}{33} \gram$ \citep{shapiro_1983}.
Such a star has a radius of approximately $5\times$ that of our own Sun, $R_\star \sim 5 R_\odot$ where $R_\odot \approx \Ten{6.96}{10} \cm$.%
\footnote{
In reality, the massive star evolves off the main sequence before the supernova explosion and can expand in its radius. 
However, the discussed values are good ``first guesses'' to perform the order-of-magnitude estimates presented in this section.}
Then, the resulting (very approximate) estimate for the mean density of the matter inside the progenitor star is $\rho_\star \approx 0.1 \gram \cm^{-3}$, roughly $\frac{1}{10}$ of the density of liquid water.

Such stars fuse lighter elements into heavier metals, all the way up to iron. 
Eventually, the heavy iron elements accumulate and sink to the bottom of the gravitational potential, forming an iron core for the star. 
This core has a theoretically known maximum mass of about $1.4 \Msun$. 
Once this limit is exceeded, the core collapses, and the outer layers explode as a bright supernova.

The remnant core consists of roughly $M/m_\mathrm{p} \sim \Msun/m_\mathrm{p} \sim 10^{57}$ atoms, where $m_\mathrm{p} \approx \Ten{1.67}{-24} \gram$ is the mass of a proton.
Suppose those atoms shed their electrons and form a giant blob of neutrons. 
In that case, the resulting remnant will have a characteristic size $\ell \sim (10^{57})^{\frac{1}{3}}  r_n \sim 10 \km$, where $r_\mathrm{n} \sim 10^{-13} \cm$ is the approximate size of an atomic nucleon. 
A compact object with a mass $\sim \Msun$ and radius $\sim 10\km$ is hence left behind--- a neutron star!
Even though these values are approximate and based on very naive conservation laws, they are not far from the actual observed values for neutron stars;
here we adopt the canonical values of a neutron star mass $M_{\mathrm{NS}} = 1.4\Msun$ and radius $R_{\mathrm{NS}} = 12\km$.

A canonical neutron star has an extraordinarily high mean density, $\bar{\rho} \sim M_{\mathrm{NS}}/R_{\mathrm{NS}}^3 \sim 10^{15} \gram \cm^{-3}$ and surface gravity, $g \sim 10^{14} \cm\second^{-2}$.
The central density of such an object, $\rho_\mathrm{c} \sim \bar{\rho}$, far exceeds the typical density of a nucleon, $\rho_\mathrm{n} \sim m_\mathrm{p}/r_\mathrm{n}^3 \sim 10^{14} \gram\cm^{-3} \ll \rho_\mathrm{c}$.
This implies that the matter is compressed so densely that there is no empty space between the atomic core and the electrons; 
instead, individual atomic nuclei are packed side by side, forming a macroscopic ``quantum ball'' of neutrons.

\subsection{First observation of a neutron star}

While the theoretical existence of neutron stars was speculated already in the 1920s, conclusive observational evidence came only at the end of the 1960s.
Neutron stars were first detected using a radio antenna in the pastures of Cambridge \citep{hewish_1968}.
The primitive antennas were originally designed for monitoring the radio emission from quasars; 
they were, however, sensitive enough to detect all kinds of human-made noise signals (arc welders, sparking thermostats, etc.)
Among many such noise signals, Jocelyn Bell, a student of Tony Hewis, noticed a signal that did not seem to have a direct connection to anything happening on Earth;
instead, the signal --- that was sometimes on, sometimes off --- seemed to originate from a specific location in the sky.
The mystery deepened when they were able to resolve the signal, using a high-speed recorder, into a regular set of pulses with a stable period of $P \approx 1.337 \second$.
The source was initially (and jokingly?) named LGM-1, standing for Little Green Men, since they did not know of any astrophysical engine capable of powering this periodic extraterrestrial signal.

By 1968, the radio observers had detected multiple pulsing signals with periods varying from $P \approx 0.25$ to $3.3 \second$. 
The sub-structure of the pulses themselves could be even shorter, $\Delta t <0.01 \second$.
The sources were also found to be extremely stable. 
There was, however, a small systematic decrease detected in many of them, on a timescale of $P/\dot{P} \sim 10^3$ to $10^6 \yr$, where $\dot{P} \equiv \ud P/ \ud t$ is the so-called period-derivative, quantifying the change of the period.%
\footnote{Such highly accurate measurements were possible because the signals were very stable and could be accumulated over long time intervals of many years.}

\subsection{Theoretical arguments for the existence of neutron stars}

So what can we conclude from these observational facts?
The following reasoning was laid out by an American astrophysicist Thomas Gold \cite{gold_1968}.

Firstly, the short duration of the pulses implies a tiny object; 
light crosses a distance of $\sim 3000\km$ in a duration of the sub-pulse, $\Delta t \sim 0.01 \second$.
This means that the emitting region must be smaller than that.
In practice, only black holes, neutron stars, or white dwarfs are small enough astrophysical objects so that their radius $R < 3000\km$.
This was the first clue: the culprit has to be some type of compact object.

Secondly, how to produce a stable pulsation from such sources?
The usual suspects that could introduce such highly periodic signals include rotation, vibration, or orbital motions.
Orbital motions are, however, ruled out because the observed periods of the signal were not increasing (corresponding to inspiral) but were found to be decreasing (i.e., negative period derivatives).
In addition, a binary system with an orbital period of $P_{\mathrm{orb}} \sim 1\second$ would merge in a few years as it loses angular momentum via gravitational radiation. 
Therefore, orbital motions could not be the source of the pulsations.

The shortest time scale out of the proposed emission mechanisms results from vibrations; here, the gravity of the object is a natural candidate for the restoring force.
The timescale for this is $t \sim 2\pi/\sqrt{G \rho}$, which, for the densities of white dwarfs, $\rho \sim 10^8 \gram \cm^{-3}$,  is $\sim 1 \second$; 
borderline too slow to explain the fastest observed pulse periods.
Conversely, for neutron stars, corresponding to characteristic densities $\rho \sim 10^{14} \gram \cm^{-3}$, the vibration timescale would be milliseconds; 
borderline too fast to explain the slowest observed pulses.
Black holes do not have a surface capable of emitting vibrations, so they do not need to be considered for this mechanism.
These conflicting timescales were used to argue against vibration as a source of the pulsations.
Furthermore, it is also quite unlikely that any surface vibration could go on for years and years without damping.

This left the contemporary researchers only with rotation.
A maximum rotation rate that a star can support before being ruptured will be close to its Keplerian frequency, $f_\mathrm{K} = (1/2\pi) \sqrt{G M/R^3}$;
it is $f_\mathrm{K} \sim 1700 \Hz$ for a neutron star and $f_\mathrm{K} \sim 0.1 \Hz$ for a white dwarf (with $M_\mathrm{WD} \sim \Msun$ and $R_\mathrm{WD} \sim 7\times10^8 \cm$).
This, in the end, helped rule out white dwarfs from the possible list of sources because they would have a hard time sticking together under such a rapid rotation.
Any surface emission from rotating black holes was also naturally ruled out because black holes do not have a surface.

This reasoning led contemporary scientists to believe that the Little Green Men observed in the Cambridge pasture originated from rotating neutron stars.

\subsection{Rotating neutron stars}

So, what rotation rates can we expect neutron stars to have?
The magnitude of the spin can be estimated from the conservation of angular momentum during the formation of the proto-neutron star.
Our own Sun rotates around itself with a period of $P_\odot \sim 25.5$ days or $\nu_\odot \approx \Ten{4.5}{-7} \Hz$.
If such a star is compressed into a neutron star, and the angular momentum is conserved during the process,%
\footnote{A realistic formation of a protoneutron star in a supernova collapse might not conserve angular momentum; 
for example, suppose there is an off-axis stream of material falling toward the center. In that case, it can result in a seeming over-accumulation of net angular momentum in the core parts.
Hence, estimates of the initial angular momentum of a proto-neutron star are very uncertain.}
the new spin would be $\nu \sim (R_\odot/R_{NS})^2 \nu_\odot \sim 2000\Hz$.
This value is a bit high, but it does get us to the right ballpark.
Observed spin frequencies of neutron stars vary from $\sim 0.01 \Hz$ up to $\sim700 \Hz$ \cite{2022NatAs...6..828C}.
Some X-ray pulsars have been argued to rotate as slowly as $\sim 10^{-5} \Hz$ \cite{2017MNRAS.469.3056S}.

The spin period of neutron stars is not constant; 
even the first pioneering observations of pulsars could detect a slow downward drift in the spin period.
This spin-down of pulsars is slow and continuous with $\dot{P}/P \sim 10^{-12} \second\second^{-1}$, and it results from magnetic dipole losses as the magnetosphere rotates and generates electromagnetic radiation which, in turn, carries a fraction of the star's angular momentum away.
This kind of spin-down is, therefore, naturally expected for isolated pulsars.
In addition, many pulsars show abrupt spin-down events, called glitches, where the spin suddenly changes in seconds.

There are also observations of neutron stars in binary systems - in fact, a big fraction of stars in any galaxy is expected to reside in systems with multiple stars;
exotic multiples up to seven stars are known \cite{2018ApJS..235....6T}.
These binary systems are evolving, and the stars commonly exchange mass.
When the mass is transferred onto the neutron star, it can be spun up: as the in-falling material streams in, hits the star off center, and exerts a torque, it can transfer some of the binary system's angular momentum into the central object.
Alternatively, the torque can be transferred to the star via the magnetic field lines that are attached to the inner disc.
This angular momentum transfer generates a sub-population of neutron stars called ``recycled millisecond pulsars'' - neutron stars spun up by accretion to spin periods of up to $P \sim 0.001 \second$.

This also highlights the importance of understanding the surroundings of neutron stars when trying to explain their physics.
In many cases, the evolution and physics of the neutron stars in binary systems are intimately influenced by the companion star.

\subsection{Magnetic fields of neutron stars}

The remaining fundamental quantity of a neutron star is its magnetic field.
The neutron star's magnetic field strength can be estimated from the conservation of magnetic flux during its birth \citep{woltjer_1964}.
A typical progenitor of a neutron star has a magnetic field of $B_\odot \sim 1\gauss$;
for comparison, a typical fridge magnet has a magnetic field of $\sim 10\gauss$.
The stellar magnetic fields are powered by dynamo processes in which the differential rotation and convective circulations inside the star amplify the magnetic field.
The field can be highly tangled and irregular due to the turbulent motions of the stellar interiors. 

When such a star collapses into a neutron star, the dynamo process is thought to halt and the internal magnetic field to be squeezed into a smaller volume. 
The resulting strength of the new field can be estimated from the conservation of the magnetic flux  through the surface ($\Phi_B = \int_S \vec{B} \cdot \ud \vec{S}$, where $\ud\vec{S}$ is an infinitesimal area element); 
alternatively, we can think that the number of magnetic field lines penetrating the star's surface is conserved.
Thus, by conservation of magnetic flux, we end up with $B_{\mathrm{NS}} \sim B_\odot (R_\odot/R_{\mathrm{NS}})^2 \sim 10^{10} \gauss$.
The resulting field is enormous; 
the largest continuously-operating human-made magnets on Earth can sustain ``only'' $\sim 10^6\gauss$.
Observed magnetic fields of regular neutron stars range from $10^8$ to $10^{12}\gauss$.

Some neutron stars are observed to house an even stronger magnetic field of $\sim 10^{14}$ to $10^{16}\gauss$---these are known as magnetars \citep{thompson_1995}.
The atypical magnetic field is thought to originate from a dynamo process that occurs during the progenitor star's collapse, amplifying the star's regular magnetic field by an unknown factor of $\sim 100$ to $1000$ to super-high strengths.
Additionally, these magnetars are thought to be so young, $t \ll 10^6\yr$, that the enormous field has not yet had time to (resistively) decay away.
These magnetars were discovered because of the most violent astrophysical events ever observed by humankind.

\subsection{Gamma-ray blasts from the past}

At 10:51 am EST on the 5th of March, 1979, two unmanned Soviet space probes, Venera 11 and Venera 12, suddenly became overloaded with an unprecedented gamma-ray shower.%
\footnote{See \url{https://solomon.as.utexas.edu/magnetar.html} for more details.}
The gamma rays first hit Venera 11 and then, 5 seconds later, Venera 12.
The onboard detectors observed a sudden jump from a regular background gamma-ray count rates of $\sim 100$ up to $\sim 40\,000$ counts per second, and then a level entirely off the scale, all during a $\Delta t \sim \mathrm{m}\second$.
After this, 11 seconds later, the gamma rays reached an American space probe Helios 2 orbiting the Sun, also knocking it out.
None of the detectors were designed to withstand such intense emission, which saturated the internal photon counters.

On that day, an unprecedented gamma-ray wavefront continued to sweep through our solar system.
After the Helios 2 probe, it reached the planet Venus, where the Pioneer Venus spacecraft detected a burst of gamma rays and quickly saturated.
After seven more seconds, the wavefront reached Earth and was detected by Vela satellite and Soviet Prognoz 7, again knocking their internal detectors out.
Even the Einstein X-ray observatory, orbiting Earth, detected a strong signal, even though it was not pointing in the direction of the blast --- the gamma rays were diffusing through its back metal shield and still overloading the detector.
A similar thing happened to the international Sun-Earth Explorer (ISEE) that was pointing in another direction, but the gamma rays penetrating through its solid metal body were still energetic enough to ramp up the detector to the maximum.

Fourteen and half hours later, another fainter burst arrived from the same spot, lasting $1.5\second$.
Then, a month later, on April 5th and again on April 24th, two more gamma-ray bursts were detected.
Humankind was registering its first giant flares from what became known as soft gamma-ray repeaters.

Tracking of the 5th of March wavefront that swept through our Solar system enabled contemporary researchers to pinpoint the origin of the blast to a nearby supernova remnant, SNR N49.
However, this supernova remnant was not even in our Galaxy but $\approx 50\,\mathrm{kpc}$ away in one of our satellite galaxies, Large Magellanic Cloud.
The intense blast had traveled almost $160\,000$ years to us and was still powerful enough to knock out our space-borne detectors.
The nature of the soft gamma-ray repeaters and these giant flares remained a mystery for years as no plausible energy source was established to explain the production of the immense gamma rays observed.

Many years later, Robert Duncan and Chris Thompson, both at Princeton University, started investigating the implications of hot and messy proto-neutron stars right after they were born in supernova explosions \cite{1992ApJ...392L...9D}.
Recent simulations seemed to indicate that the dense fluid in the core had strong convective motions that helped to carry the heat out from the hot remnant.
The ultra-dense matter at the interior would also carry electric currents; hence, a dynamo process similar to Earth's core could, in theory, be possible.
If the neutron star would collapse fast enough, rotate rapidly, and have strong convective motions, it could house a magnetic dynamo inside of it, which could amplify any existing magnetic field by twisting and folding it to anywhere from $10$ to $1000$ times of its original strength.
Therefore, if typical pulsars with $B \sim 10^{12}\gauss$ were the outcome of a failed dynamo, what would happen if the dynamo succeeded?
They hypothesized that a successful dynamo process would result in a neutron star with an enormous magnetic field up to $B \sim 10^{16}\gauss$ field.
These monsters they, later on, dubbed ``magnetars''.

The first observational implication of such a field would be that if the star's internal stresses twisted and shook the magnetosphere, it could exhibit spectacular flares similar to our own Sun.
Therefore, in the case of a magnetar, the flare could turn a fraction of that immense magnetic field into radiation, and, in theory, power intense blasts of gamma-rays with luminosities $L \gtrsim 10^{43} \ergs$, capable of explaining the observed giant flares saturating our X-ray and gamma-ray detectors.
Eventually, this became the accepted explanation for the giant flares, rendering magnetars the kings of the magnetic fields in the known Universe.

\subsection{Many observational faces of neutron stars}

As we have seen, neutron stars have multiple observational "faces".
We have detected both isolated neutron stars and neutron stars in binary systems.
The types of companion stars in the binary systems vary from low-mass stars to other neutron stars (or even black holes).
Some neutron stars are powered by rotation (pulsars), and some only by their internal heat (e.g., central compact objects found in supernova remnants).

There is a distinct dichotomy in the observed spin distribution:
regular pulsars, identified typically as relatively young neutron stars, are slowly spinning ($P \sim 1\second$), whereas older neutron stars in binary systems have been spun up by accretion, giving rise to what we know as recycled millisecond pulsars.
These recycled millisecond pulsars can be further divided into rotation-powered (i.e., radio millisecond pulsars, thought to operate similarly to regular pulsars but to spin faster) and accretion-powered (i.e., pulsating on X-ray frequencies due to the in-falling material that is channeled to the magnetic poles, creating hot spots onto the surface) sources.
A few stars, known as transitional millisecond pulsars, are observed to transition between these two states.

Some of the rotation-powered millisecond pulsars are also identified as ``spiders'' (named so because these stars are acting nasty on their companion stars).
Spiders generally have irregular radio eclipses due to intra-binary material driven off the donor by a pulsar wind.
They can be further divided into black widows (characterized by very low-mass companions of $M_c < 0.06 \Msun$) and redbacks (characterized by hydrogen-rich companions of $M_c \gtrsim 0.1 \Msun$).

Various classes of neutron stars and their observational phenomena are listed in Table~\ref{tab:types}.

\newgeometry{margin=1cm} 
\begin{landscape}

\begin{table}[ht]
\caption{Different observational classes of neutron stars}
\begin{center}
 \begin{tabular}{ p{0.05\linewidth} | p{0.5\linewidth}| p{0.04\linewidth} | p{0.3\linewidth}}
    \textbf{Class} & \textbf{Description} & \# &  \textbf{Observed phenomena}\\[3ex]
    \multicolumn{3}{c}{\textbf{Isolated neutron stars}}\\[1.5ex]
    CCOs & Weakly-magnetized young neutron stars in centers of supernova remnants & $\approx 10$ $^a$ & Constant thermal-like soft X-ray emission. \\[1.5ex]
\hline
    Pulsars & Spinning, magnetized, young neutron stars & 3320 $^b$ & Pulsating radio emission, pulsating X-rays, gamma-ray emission. \\[1.5ex]
\hline
Magnetars & Very young neutron stars with extremely strong magnetic fields. Divided into soft gamma repeaters (SGRs) and anomalous X-ray pulsars (AXPs). & 30 $^c$ & SGRs: hard X-ray flares, giant bursts. AXPs: persistent and pulsating X-ray emission, transient radio emission. Additionally: fast radio bursts. \\[3ex]
        \multicolumn{3}{c}{\textbf{Low-Mass X-ray Binary systems ($M_c \lesssim \Msun$)}}\\[1.5ex]
    Bursters & Accreting neutron stars, which depict thermonuclear X-ray bursts & $118$ $^d$ & Type-I thermonuclear X-ray bursts, burst oscillations. \\[2.5ex]
    \multicolumn{3}{c}{\textit{Recycled millisecond pulsars}}\\[1.5ex]
 RMSPs & Rotation-powered millisecond pulsars. Like regular pulsars but spun up by angular momentum gain from the accretion. & 472 $^e$ & Millisecond-period radio pulsations, X-ray and gamma-ray emission. \\[1.5ex]
 \hline
 AMSP & Accretion-powered millisecond pulsars. & 18 $^e$ & X-ray pulsations. Sometimes also thermonuclear bursts.\\[1.5ex]
\hline
TrMSPs & Transitional millisecond pulsars. MSPs transitioning between accretion-powered and rotation-powered states & 3 $^f$ & X-ray pulsations, X-ray flaring, transient optical emission.\\[3ex]
    \multicolumn{3}{c}{\textit{Spider binaries}}\\[1.5ex]
    Black widows & RMSPs with irregular radio pulsations. Very low companion masses of $M_c < 0.06\Msun$ & 38 $^e$ & X-ray emission, radio emission, optical measurements of the companion. \\[1.5ex]
\hline
    Redbacks & RMSPs with irregular radio pulsations. Hydrogen-rich companion with $M_c \gtrsim 0.1\Msun$ & 22 $^e$ & X-ray emission, radio emission, optical measurements of the companion. \\[2.5ex]
        \multicolumn{3}{c}{\textbf{Other Binary systems}}\\[1.5ex]
\hline
    HMXBs & Accreting neutron stars with high-mass ($M_c \gtrsim 10 \Msun$) companions & \cite{2022arXiv220414185M} & constant X-ray emission, X-ray pulsations \\[1.5ex]
\hline
     Merging compact binaries & Coalescing binary systems composed of a neutron star and another compact object (NS-WD, NS-NS, or NS-BH). These systems can also be observed when they coalesce and merge. & 4 $^g$ & precursor emission (?), gravitational waves, short gamma-ray bursts, kilonova (seen in IR, optical, UV).  \\[3ex]
    \multicolumn{3}{c}{\textbf{Environments}}\\[1.5ex]
    PWNs & Pulsar wind nebulae. The region inside a supernova remnant powered by a wind from a central pulsar. & $33$~+~$28$~$^h$ & Broadband emission from radio to X-rays and gamma-rays. Cosmic rays.\\[1.5ex]
    \hline
    SNRs & Supernova remnants. Large-scale structures that are the result of supernova explosions. The region is bounded by an expanding shock wave. & $\approx 294$~$^i$ & Multi-band electromagnetic emission, cosmic rays.\\[1.5ex]
\hline
\end{tabular}
\end{center}
\label{tab:types}
\begin{center}
\textbf{Notes:}\\ 
$^a$ \url{http://www.iasf-milano.inaf.it/~deluca/cco/main.htm} (list of observed CCOs);\\
$^b$ \url{https://www.atnf.csiro.au/research/pulsar/psrcat/} (pulsars catalog; \cite{2005yCat.7245....0M});\\
$^c$ \url{http://www.physics.mcgill.ca/~pulsar/magnetar/main.html} (list of observed magnetars);\\
$^d$ \url{https://personal.sron.nl/~jeanz/bursterlist.html} (list of observed Type-I bursters);\\
$^e$ \url{https://blacksidus.com/millisecond-pulsar-catalogue/} (millisecond pulsar catalog);\\
$^f$ Patruno et al. 2022 \cite{2022ASSL..465..157P}; \\
$^g$ GW170817, GW190425, GW200105, GW200115;\\
$^h$ PWN $+$ PWN w/o coinciding pulsar \url{http://www.physics.mcgill.ca/~pulsar/pwncat.html} (PWN catalog;  \cite{2006csxs.book..279K});\\
$^i$ \url{https://www.mrao.cam.ac.uk/surveys/snrs/snrs.data.html} (list of SNRs;  \cite{2019JApA...40...36G}); \\
\end{center}
\end{table}
\end{landscape}
\restoregeometry

\newpage
\section{Laboratories of gravitation}

The extreme density of the matter that is concentrated inside neutron stars makes them ideal laboratories of gravitation at the extreme limit because they can strongly deform the spacetime metric \citep{misner_1973, shapiro_1983, paschilidis_2017}:
the high compactness of the star will curve the spacetime, whereas the rapid rotation can twist it.
These effects create complex spacetime metric deformations, which modify the visual appearance of neutron stars. 
Detailed modeling of these effects allows us to estimate the radius and mass of neutron stars from observations.

Another frontier in the gravitational physics of neutron stars is the gravitational wave observations from merging neutron star binaries \citep{baiotti_2019}.
Modeling these systems relies on understanding and exploiting the dynamical, time-dependent side of gravity: as the coalescing, space-time-deforming neutron stars merge, they will simultaneously vibrate spacetime itself.
The resulting gravitational waves can then be detected and compared to general relativistic hydrodynamic simulations to constrain the properties of the neutron star matter.
In practice, this allows measuring the tidal deformability of the ultra-dense matter of the coalescing neutron stars.

\subsection{Space-time deformations}

The high density of the neutron star leads to strong space-time distortions.
The exterior space-time metric of a static, non-rotating spherical object is given by the Schwarzschild metric \citep{misner_1973}%
\footnote{We follow the \citealt{misner_1973} sign convention and use a metric signature of ($-,+,+,+$).},
\begin{equation}
    \ud s^2 = -(1-u) \ud t^2 + \frac{\ud r^2}{1-u} + r^2 \ud\theta^2 + r^2 \sin^2\theta \ud\phi^2,
\end{equation}
where $t$ is the coordinate time, 
$r$ is the radial coordinate (defined so that the area at a time fixed time is $4\pi r^2$),
$\theta$ is the latitudinal angle,
$\phi$ is the longitudinal angle, and
$u$ is the compactness parameter,
\begin{equation}
    u \equiv \frac{R_\mathrm{S}}{R_{\mathrm{NS}}} = \frac{2 G M_{\mathrm{NS}}}{R_{\mathrm{NS}} c^2},
\end{equation}
$G$ is the gravitational constant, and 
$R_\mathrm{S} \approx 2.95 (M/\Msun)~\km$ is the Schwarzschild radius.%
\footnote{Sometimes it can be more convenient to use the isotropic Schwarzschild metric of the form 
    \begin{equation}
\ud s^2 = -\left( \frac{1-\frac{\ub}{2}}{1+\frac{\ub}{2}} \right)^2 \ud t^2 + \left( 1+\frac{\ub}{2} \right)^4(\ud\rb^2 + \rb^2(\ud\theta^2+\sin^2\theta \ud\phi^2)),
    \end{equation}
    where $\bar{r}$ is the isotropic radial coordinate and $\bar{u} \equiv M/\bar{r}$ is the isotropic compactness.
    Surfaces of constant time in this metric are conformally flat; therefore, all the angles are represented without distortion.
    However, in this case, the measured distances depend on the location.
    See \citep{nattila_2018}.
}
A neutron star is so compact that the corresponding Schwarzschild radius, $R_\mathrm{S} \approx 4.1\km$, is slightly smaller than the star's physical radius, $R_\mathrm{NS}/R_\mathrm{S} \approx 3$.
Hence, the star is on the brink of collapsing into a black hole.

The neutron star compactness parameter, $u$, measures the severity of space-time distortions.
Its value is somewhere between $u \approx 0.3 - 0.5$, leading to strong dilation effects of the coordinate time $t$ ($\propto \sqrt{1-u} \approx 0.8$) and stretching of the radial coordinate ($\propto 1/\sqrt{1-u} \approx 1.2$) of the space-time close to the star's surface.
Escape velocity from on top of such an object is very large, $v_\mathrm{esc} = \sqrt{2 G M_\mathrm{NS}/R_\mathrm{NS}} \sim \Ten{1.93}{10} \cm \second^{-1}$, and is comparable to the speed of light, $\sim c/2$.

\subsection{Rotating stars}

Rotation of the compact star, and hence the large angular momentum, will lead to additional shear-like distortions of the space-time metric \citep{paschilidis_2017}.
In this case, we need to specify the exact latitude where the effect is studied;
a typical choice is an equator with the corresponding equatorial radius $R_\mathrm{e}$, defined such that the circumference of a star (using the Schwarzschild metric radial coordinate) is $4\pi R_e$, as measured in the local static frame. 
The radius at the pole is smaller, $R_\mathrm{p} < R_{\mathrm{e}}$.

For a star rotating with an angular velocity $\Omega$, the moment of inertia is $I_\mathrm{NS} \sim 10^{45} \gram \cm^2$.
Angular momentum is then $\mathcal{J} = I \Omega$.
The dimensionless angular velocity is
\begin{equation}
    \bar{\Omega} \equiv \frac{\Omega}{\Omega_\mathrm{K}} =\sqrt{ \frac{\Omega^2 R_\mathrm{e}^3}{G M_\mathrm{NS}} },
\end{equation}
where the Newtonian mass shedding limit (Kepler limit) is $\Omega_\mathrm{K} = \sqrt{G M/R_\mathrm{e}^3}$.
Observed neutron stars have $\bar{\Omega} \sim 0.01 - 0.1$ resulting in strong rotational spacetime effects.

Rotation can be incorporated into the space-time metric by expanding around the small parameter $\bar{\Omega}$. 
The first-order expansion corresponds formally to the Kerr-like metric, where an extra angular velocity term of a local inertial frame appears.
This is known as the frame-dragging effect, where the spacetime is rotating together with the star (i.e., in practice, velocity $v \rightarrow (v - v_\mathrm{fd})$, where $v_{\mathrm{fd}}$ is the frame-dragging velocity).
Second-order expansion is formally related to Hartle-Thorne-like metric \cite{hartle_1968}, which, in addition to the frame-dragging effects, has (two) quadrupole moments entering into the metric description.%
\footnote{These are the multipole moments of the energy density and pressure.
    Their value depends on the coordinate system, whereas a single coordinate-invariant quadrupole moment can be obtained as a combination of the two \citep{pappas_2012}.
}
\citep{braje_2000, cadeau_2007, baubock_2013, algendy_2014, nattila_2018}

The rotation will also modify the actual shape of the star, making it oblate--- the rotating star develops a bulge on the equator.
The shape is numerically found to be close to an oblate spheroid, \citep{morsink_2007, algendy_2014, nattila_2018, suleimanov_2020},
\begin{equation}
R(\theta) \approx R_e \left[ 1- \left(\frac{\Omega}{\bar{\Omega}} \right)^2 (0.788 - 2.06 u) \cos^2\theta \right],
\end{equation}
where $R(\theta)$ is the radius at a colatitude $\theta$ (angle from the pole towards the equator).

\subsection{Radiation from the star's surface}


The strong stretching of the metric in the radial direction modifies the apparent visual look of neutron stars for distant observers.
The most prominent effect is the gravitational light bending around the star that increases the apparent radius (that we would measure by looking at the star from a distance) to be 
\begin{equation}
    R_\infty = \frac{R}{\sqrt{1-u}}.
\end{equation}

Similarly, the dilation of the time component modifies the energy of the emanating photons.
This effect, gravitational redshift, is also connected to $u$, and is given as
\begin{equation}
    1 + z = \frac{1}{\sqrt{1-u}}.
\end{equation}
It is a measure of how much, for example, the frequency changes, $\nu_f/\nu_i = (1+z)^{-1}$ (where $\nu_f$ is the observed and $\nu_i$ initial photon frequency), as a photon climbs out from the gravitational well created by the neutron star.
The typical effect is a redshift, $\nu_i > \nu_f$, where the frequency is reduced during the climb.

\begin{figure}[t]
\centering
\includegraphics[clip, trim=3.0cm 1.0cm 2.0cm 2.0cm, width=5cm]{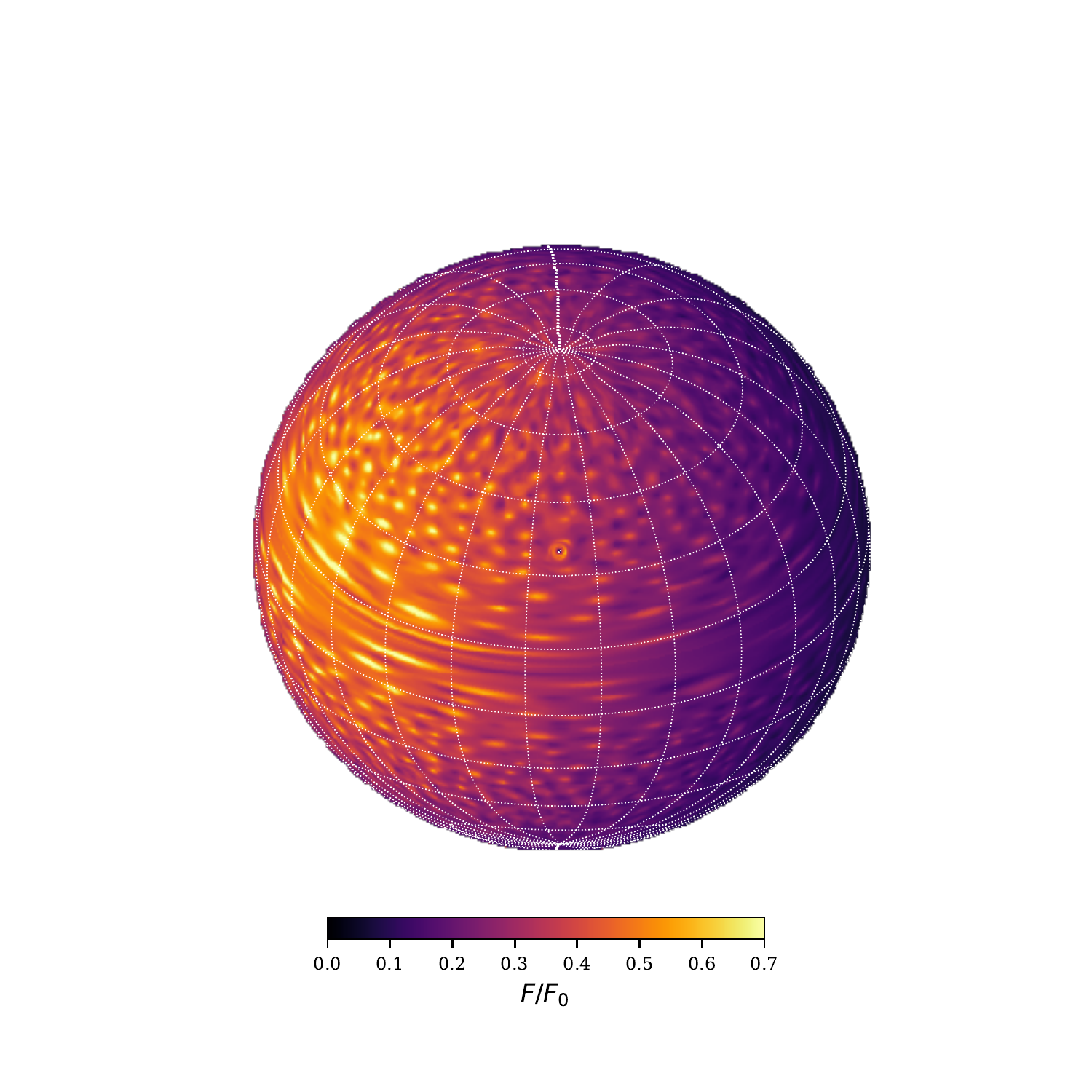}
\includegraphics[clip, trim=3.0cm 1.0cm 2.0cm 2.0cm, width=5cm]{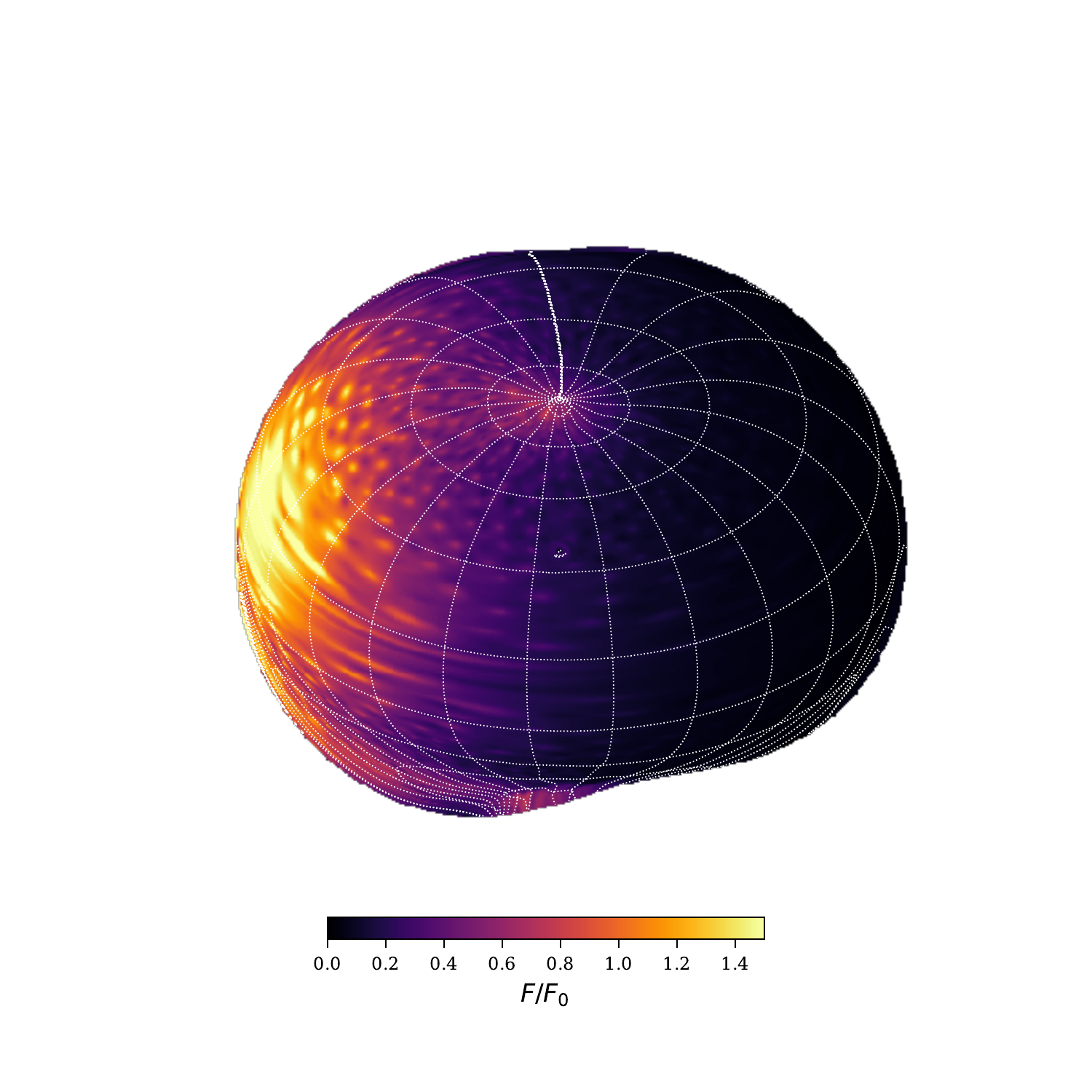}
\caption{\label{fig:boosting}
    Example images of rapidly rotating neutron stars.
    The colormap shows the radiation flux, $F \propto (1+z)^{-4}$, from rotationally distorted neutron stars as a function of the non-red-shifted, non-boosted flux, $F_0$ ($z = 0$).
    Left panel shows an image of a typical massive neutron star ($M_{NS} = 2\Msun$, $R_{NS} = 13\km$) spinning with a frequency $f = 600 \Hz$ being observed with an inclination angle (w.r.t. to the spin axis) of $i = 60^\circ$.
    Right panel shows the visual appearance of the same neutron star with an extreme spin frequency of $f = 1400 \Hz$ close to the mass-shedding limit.
    The ray-tracing is performed with the general relativistic ray-tracing code Arcmancer, \citep{pihajoki_2019}.
  }
\end{figure}

The photon energy is also modified by the Doppler-like boosting factor,
\begin{equation}
    \delta = \frac{\sqrt{1 - \beta^2}}{ 1- \beta \cos\zeta},
\end{equation}
where 
\begin{equation}
    \beta(\theta) = \frac{R \Omega}{c \sqrt{1-u}} \sin\theta,
\end{equation}
is the surface velocity of the rotating star,
and $\zeta$ is the angle between the photon momentum and the surface velocity vector, $\vec{\beta} = \beta(-\sin\phi, \cos(\phi), 0)$ at the emission point.
The Doppler boosting of the radiation can lead to a blue shifting of the radiation, $\nu_f > \nu_i$, causing an increase in the frequency, as photons are ``slingshot'' away from the surface like being thrown out from a rapidly rotating cartwheel.

The total frequency (or energy) change is the product of the gravitational and Doppler-boosting factors, $\nu_f/\nu_i = \delta/(1+z)$.
Observed flux scales as $F_E \propto (\nu_f/\nu_i)^3$ and energy-integrated flux as $F \propto (\nu_f/\nu_i)^4$.
The emergent flux is visualized in Fig.~\ref{fig:boosting} for two different example cases.
Finally, we note that all the radiative processes should be defined in the co-rotating frame of the star (i.e., for an observer located at the surface).

To summarize, the emission calculations typically take into account space-time deformation effects up to second-order in rotation, which are composed of:
\begin{enumerate}[label=\roman*]
    \item gravitational potential (and hence gravitational redshift),
    \item Doppler-boosting, which originates from the rotating surface,
    \item change in the local emission angle (affecting the limb-darkening and brightening effects), 
    \item time transformations (because the integration of finite-sized areas needs to take place simultaneously), and
    \item oblate shape of the star.
\end{enumerate}
To a lesser extent, we also need to take into account the following:
\begin{enumerate}[label=\roman*, resume]
    \item quadrupole-deformations of the spacetime (modifying the gravitational redshift term),
    \item frame-dragging effect (reducing the angular velocity close to the equator), and
    \item quadrupole deformations which modify the differential emission area element ($\ud S = R^2 \ud\cos\theta \ud\phi$ for a flat spacetime metric).
\end{enumerate}

Typical neutron star images are constructed using the so-called S$+$D approximation (Schwarzschild metric and Doppler boosting) \citep{pechenick_1983, poutanen_2003, poutanen_2006}, where the observer's polar coordinates are connected to the star's co-rotating spherical coordinates (algebraically) and quantities computed at the surface of the star are Lorentz-boosted into the static non-rotating observer's frame.
The method can be refined by taking into account the oblate shape of the star \citep{cadeau_2007, morsink_2007, lo_2013, nattila_2018, salmi_2018, bogdanov_2019, loktev_2020}.

\subsection{Pulse profile modeling}

If the surface of the neutron star has a spot that is located off the rotational axis (in comparison to the whole surface just emitting uniformly), it produces periodic pulsations with the spin frequency of the star.
Information in these pulse profiles can be used to constrain the geometry of the rotating system together with the mass and radius of the neutron star.

Creation of such spots requires some physical effect that breaks the axisymmetry (along the azimuthal direction) on the surface and makes an azimuthal patch of the surface brighter (or, in theory, dimmer, corresponding to cold spots that produce flux deficit as a function of the phase).
Three mechanisms for spot creation are typically considered:
heated magnetic polar caps in isolated pulsars, accretion-heated spots in binary systems, and dynamic thermonuclear burning regions.

\textbf{Heated magnetic polar caps of isolated pulsars.}
In isolated pulsars, the magnetospheric activity above the surface leads to acceleration of charged particles that are launched towards the surface \cite[e.g.,][]{2007Ap&SS.308..419R}.
The in-falling ``rain'' of charged particles leads to heating the star's surface- a hot spot.
The opposite magnetic pole of the star can house a second antipodal spot.
There are observational implications that some of such polar caps might not be circular but resemble more arc-like structures (banana-shaped spots) \cite{2019ApJ...887L..21R,2019ApJ...887L..24M}.
Such complex structures can be created, for example, if the magnetic field of the neutron star is not a dipole but has higher-order multipole components due to a more tangled internal field.

\textbf{Accretion-heated spots of neutron stars in (low-mass X-ray) binary systems.}
Accretion-heated spots are generally labeled as originating from accreting millisecond pulsars.
In these systems, the surface is heated by the material from the companion star, which falls into the neutron star.
The spot is formed as the magnetic field of the neutron star channels the material into a stream that hits the surface and heats it.

\textbf{Thermonuclear burning fronts.}
The thermonuclear burning of accumulated material on top of accreting neutron stars (in low-mass X-ray binary systems) can sustain a coherent burning region, similar to hurricanes on Earth's weather layer. 
The dynamics of these storms are still not very well understood: 
The burning fronts can lead to spots that drift (i.e., the spot is slowly moving on the surface) and expand (i.e., the burning front propagates into the fresh fuel), making their modeling very complicated.

\phantom{xx}

Irrespective of the origin of the spots, the information in the observed pulsating emission --- pulse profile --- can be used to measure the radius and mass of the neutron star housing the spot.
The simplest assumption for an emitting spot is a circular emission region at a co-latitude $\theta_\mathrm{s}$ with a half-angular size $\rho$ (measured in degrees using great-circle distances).
During one rotation of the star, the spot first becomes visible to the observer, then crosses the visible surface, and finally disappears behind the star.
This leads to a phase-resolved pulse profile where the detected emission increases, peaks, and decays.
Depending on the observer's inclination angle $i$ and on the spot's colatitude, the spot can disappear behind the star, experience a partial occultation, or always remain visible.

As an example, NASA's NICER (Neutron Star Interior Composition Explorer) mission has used the pulse profile analysis to constrain the radius and mass of the rotation-powered millisecond pulsar PSR J0030$+$0451 to be $R \approx 12.7\pm 1.2\km$ and $M \approx 1.34 \pm 0.16 \Msun$, as measured by \cite{2019ApJ...887L..21R}; or,
$R \approx 13.0^{+1.3}_{-1.1} \km$ and $M\approx 1.44 \pm 0.15 \Msun$, as measured by \cite{2019ApJ...887L..24M}.
These NICER observations also indicated that emission areas were, in fact, more consistent with three elongated arcs instead of just two antipodal circular spots.
This might imply that the magnetic field structure is more complex than the regularly assumed dipole field.
Similarly, the NICER collaboration has constrained the radius of a massive pulsar PSR J0740$+$6620 with a known mass of $M \approx 2.08 \pm 0.07 \Msun$ to be $R \approx 12.4^{+1.3}_{-1.0} \km$, as measured by \cite{2021ApJ...918L..27R}; or,
$R \approx 13.7^{+2.6}_{-1.5} \km$, as measured by \cite{2021ApJ...918L..28M}.
For comparison, a similar pulse profile analysis was performed with RXTE (Rossi X-ray Timing Explorer) data to constrain the radius of the accretion-powered millisecond pulsar SAX J1808.4$-$3658 to be $R \approx 11.9\pm 0.5\km$ (assuming mass $M = 1.7 \Msun$) \cite{salmi_2018}.
These measurements agree with neutron stars with a radius of $R \approx 12\km$.

\subsection{Gravitational waves}

So far, we have discussed static gravitational effects like the gravitational bending of light rays or the redshift of photon energy. 
Recent pioneering observations of gravitational waves from neutron star mergers are a good example of dynamic gravitational phenomena where spacetime changes as a function of time \cite{baiotti_2019}.

A typical toy picture of a gravitational wave depicts the wave on a vibrating elastic membrane with a surface wave-like perturbation moving on it.
This conforms to the geometric interpretation of gravity as a spacetime curvature, i.e., general relativity.
Alternatively, we can imagine the gravitational waves as a consequence of special relativity in which a change in the gravitational field takes time to be felt at a distance;
any perturbation (even of spacetime) will propagate with a maximum velocity of $c$.

Imagine a general field with strength decreasing as an inverse square of the distance, $\propto r^{-2}$.
Such a field can be visualized by a familiar picture of radial field lines originating from the source of the field.
If such a source is uniformly moving, the field lines will still have the same radial form.
However, if the source is accelerating, there will be a propagating ``dividing zone'' between a moving radial field line region close to the source (expanding with a maximum velocity of $c$) and a further-away exterior region still depicting a field of a non-moving source.
The discontinuous zone dividing the interior and exterior regions has a width $c \Delta t$, where $\Delta t$ is the duration of the acceleration.
The discontinuity can be understood as a wave traveling away from the accelerating source.
This forms an alternative basis for understanding the nature of gravitational waves.

Mathematically, the simplest object that can launch such a pulse is a time-varying monopole, $\dot{X} \equiv \ud X/\ud t \ne 0$, where $X$ is, for example, the electric charge ($X \rightarrow q$, where $q$ is an electric charge; relevant for theories of electromagnetism) or mass ($X \rightarrow M$, where $M$ is mass; appropriate for theories of gravitation).
However, the conservation of electric charge and conservation of mass prevents the existence of time-varying monopoles. So any perturbation traveling on a field generated by any kind of ``change of a monopole'' is forbidden.

Let us next consider dipoles, the second-most simple mathematical object in line.
In electromagnetism, an electric dipole (i.e., a pair of balanced positive and negative charges) is given as $P_q = \sum q_i s_i$, where $s_i$ is the distance of the charge from the center of the charge cloud.
Acceleration (for example, a rotating motion) corresponds to a non-zero second-time derivative: 
$\ddot{P_q} \equiv \partial^2 P_q/\partial t^2 = \sum q_i a_i \ne 0$, where $a_i \ne 0$ is the acceleration of the $i$th charge.
As is well known, this leads to electromagnetic dipole radiation.
We return to this form of energy loss in the electromagnetism section.

Gravitation turns out to be more complex since momentum is always conserved (in a closed system), and so $\ddot{P_M} \equiv \partial^2 P_M/\partial t^2 = \sum M_i a_i = 0$.
To make a system emit gravitational radiation, time-varying higher-order moments are required.
The subsequent mathematical entity in the multipole expansion after the dipole moment is the quadrupole moment (e.g., four balanced charges arranged in the corners of a square).
The quadrupole moment of mass distribution is defined as $I_M = \sum M_i s_i \otimes s_i$, where $\otimes$ denotes the tensor product of two vectors, resulting in a quadrupole moment being a rank-2 tensor that is described with a $3 \times 3$ matrix.
The corresponding physical quantity is the gravitational tidal field, $g' = \Delta g/\Delta d$, which represents an observable relative acceleration (i.e., a force) between two displaced particles within a small displacement of $\Delta d$ (i.e., the gradient of gravity).

The dominant gravitational radiation originates from the fourth time-derivative of the quadrupole moment, $\ddddot{I}$, which produces a change in the gravitational (tidal) field $g$ per displacement $s$ of $g' = \partial g/\partial s$ ($\sim \Delta g/\Delta d$, where $\partial g \sim \Delta g$ and $s \sim \Delta d$).
This change is proportional to $g' \sim \ddddot{I}/r$, where $r$ is the distance from the source.
Other, weaker modes scale as $g' \sim I/r^5$, $\dot{I}/r^4$, $\ddot{I}/r^3$, and $\dddot{I}/r^2$; 
all falling off more rapidly than the dominant $\ddddot{I}/r$ term.
The dominating mode contains terms of the form $M s \ddot{a}$, $M v \dot{a}$, and $Ma^2$;
for oscillatory motion, relevant for rotating systems with equal masses $M$ moving over distance $s$ with frequency $f$, all of these scales as $\propto M f^4 s^2$.
The amplitude of the gravitational radiation then scales as 
\begin{equation}
 g' \sim \frac{G M }{c^4} \frac{f^4 s^2}{r} \sim G M \frac{s^2}{\lambda^4} \frac{1}{r},
\end{equation}
where $\lambda$ is the wavelength.

In gravitational wave physics, the most commonly used parameter characterizing the amplitude of the wave is not $g'$ (i.e., change in gravity per displacement) but the dimensionless strain, 
\begin{equation}
    h \equiv 2 \int \int g' \ud t^2 \sim 2 \frac{\Delta [\Delta d]}{\Delta d},
\end{equation}
(i.e., change in displacement per displacement).
The two integrals in the strain formula represent an instantaneous change in displacement as a function of time.
Therefore, $h$ is twice the fractional change in the displacement between two nearby masses due to the gravitational waves.
This can be understood geometrically as follows.
Similar to electromagnetic radiation, this displacement is on a plane transverse to the direction of the motion of the waves;
i.e., the wave is a shear wave.
It stretches the spacetime along one axis and squeezes the other axis orthogonal to it. 
The net distortion is twice as large as what the stretching or squeezing would independently give.
Note also that $h$ itself is not observable; only the second derivatives and higher of $h$ (that produce acceleration) are detectable.

The final scaling of the strain is 
\begin{equation}
h \sim \frac{G M}{c^2} \frac{1}{r} \left(\frac{v}{c}\right)^2.
\end{equation}
In the most standard, already observed scenario of two in-spiraling neutron stars, the rotating binary system corresponds to a non-zero quadrupole moment that acts as a cyclic source of gravitational waves.
A neutron star with a mass $M \sim 2 \Msun$, inspiral velocity $v \sim c$ (just before the merger), at a distance $r \sim 6 \times 10^{25} \cm$ ($\approx 20$ megaparsecs) would result in a peak dimensionless strain of $10^{-20}$.
This kind of fluctuation in the strain was indeed observed in the GW170817 event \cite{2017PhRvL.119p1101A}.

Other possible mechanisms for generating gravitational  wave radiation from neutron stars are, e.g., breaking of axisymmetry in rotating pulsars; 
here a small mountain on top of the neutron star (corresponding to some defect of the crust) would also lead to non-zero gravitational quadrupole moment, and hence emission of gravitational waves.
The perturbation amplitude from rotating mountains is much weaker, and these signals have not yet been detected.

\subsection{Interpretation of gravitational waveforms}

Latest observations of gravitational waves from merging double neutron star binary systems offer a powerful way for constraining the properties of the ultra-dense matter inside the stars \cite{baiotti_2019}.
The gravitational wave emission observed from neutron stars is often split into the chirp and ringdown phases.
The chirp phase occurs at the beginning of the inspiral as the two stars begin to coalesce toward each other but are not yet touching.
In this case, the gravitational wave frequency is just twice the orbital frequency of the binary, $f_{\mathrm{GW}} \approx 2 f_\mathrm{K}$, where $f_\mathrm{K}$ is the Keplerian orbital frequency.
The contact frequency, the frequency reached at the contact between the two objects, is well-approximated as 
\begin{equation}
    f_c = \frac{c^3}{2 G} \frac{ \bar{c}^{3/2} }{M},
\end{equation}
where $\bar{c} = G M / R c^2$ is the average compactness, assuming identical stars with $M_1 = M_2 = M$ and $R_1 = R_2 = R$.
The chirp frequency is observable for tens of seconds and peaks at $f_\mathrm{GW} \sim 1 \,\mathrm{kHz}$.
During the chirp phase, the equation of state can be probed by measuring the dimensionless quadrupole tidal deformability (polarizability coefficient),
\begin{equation}
\Lambda = \frac{2}{3} k_2 \left( \frac{R c^2}{G M}\right)^5,
\end{equation}
where $k_2$ is the quadrupole Love number.
Its measured posterior distribution is currently consistent with $\Lambda = 0$ (corresponding to two black holes merging).
However, since the system is a known double neutron star binary (and, in addition, because a gamma-ray burst was detected), we are confident that the GW170817 event measured the tidal deformability of the ultra-dense matter inside neutron stars.
This measurement rules out the softest equations of states that require large $\Lambda \gg 0$.

As the two stars merge, they will either collapse directly into a black hole or remain, for a while, as a ``joint'' neutron star.
The resulting star can be either a stable neutron star supported by the standard pressure of the matter or a quasi-stable supra massive/hyper massive neutron star, supported (at least partially) by the centrifugal acceleration from the rotation of the object.
Supramassive neutron star remnants assume a uniform rotation profile, $\Omega(r) \propto r$ (where $r$ is the radius from the center of the remnant), whereas hypermassive remnants have differentially rotating interiors.

The cores of the two merging stars continue to rotate inside a common envelope for a while, resulting in a ring-down gravitational wave emission.
The ringdown phase is short (the system loses angular momentum rapidly due to efficient dissipation caused by the dense surrounding cloud) and emits gravitational waves at high frequencies.
The ringdown phase has three characteristic frequencies, $\Omega_1 < \Omega_2 < \Omega_3$, with a distinct physical origin.
The physical origin of the frequencies can be understood by a toy model describing a rotation of two balls connected by a spring.
The balls mimic the cores of the initial neutron stars, which rotate in a common envelope.
Such a system can have three characteristic configurations if the system is rotating and the spring is vibrating.
When the balls are at their maximum separation, the system has the largest moment of inertia and, conversely, the smallest rotation frequency, $\Omega_1$.
At their closest separation, the system has the smallest moment of inertia and, conversely, the highest rotation frequency, $\Omega_3$.
If there is no dissipation, the angular frequency of the system will oscillate between these two states.
Time spent at a given frequency	is $\propto \Omega/\dot{\Omega}$, and so most time is spent at $\dot{\Omega} \approx 0$, i.e., at the two extreme states, resulting in power-spectrum peaks at frequencies corresponding to $\Omega_1$ and $\Omega_3$.
If the spring is dissipative, the system asymptotes to an average frequency, $\Omega_2 \approx (\Omega_1 + \Omega_3)/2$.
This leads to a time-growing third peak at a frequency of $\Omega_2$.

Realistic merger simulations confirm that the initial few oscillations of the ringdown show a gravitational wave spectrum with two frequency peaks at $f = \Omega_1/\pi$ and $\Omega_3/\pi$.
Realistic merger remnant configurations are highly dissipative, so they quickly develop a strong peak at $f \approx \Omega_2/\pi$, with weaker sidebands at $f_1$ and $f_3$.
Especially the $f_1$ frequency correlates with the compactness, $u \equiv 2 G M/R c^2$, enabling a more accurate equation of state constraints in the future when observations of the ringdown phase become possible.

Finally, we note that the (short) gamma-ray bursts occurring shortly after the neutron star mergers are also an important new avenue of research \cite{2016ARNPS..66...23F}.
They are sources of electromagnetic radiation (so-called kilonova) and possible neutrinos.
They are also speculated to be responsible for the heavy element production in our Universe since the merger fuses extremely heavy elements, and the gamma-ray burst expels them to the surrounding interstellar medium.

\newpage
\section{Laboratories of nuclear physics}

The density inside neutron stars is so high, and the particles are packed together so tightly that quantum effects play an essential role in controlling the behavior of the matter \citep{haensel_2007}.
This makes neutron stars exciting laboratories of nuclear physics.
Furthermore, since the interiors of neutron stars are the densest form of observable matter known in the Universe, they provide a pathway into new studies exploring particle physics and quantum chromodynamics --- something previously thought possible only with particle colliders.
This line of research of the neutron star interiors is encapsulated into the unknown equation of state of the matter, $P(\rho, T, \ldots)$, describing a pressure $P$ as a function of thermodynamic properties of the matter like density, $\rho$, temperature, $T$, and so on.

\subsection{Dense matter inside compact objects}

When the matter is compressed from ordinary matter towards the ultra-dense states inside neutron stars, different forces play a role hierarchically.
First, the pressure is supported by the thermal motions of the gas. 
When compressed further, the charge repulsion between negatively charged electrons and positively charged protons becomes important.
Both of these effects are explainable by classical physics and are used to describe the matter and equation of state inside regular stars.

For compact objects, like neutron stars and white dwarfs, quantum mechanics is involved when explaining the behavior of matter.
First, for dense enough matter (such as those found inside white dwarfs or outer layers of neutron stars), the electrons start to be packed so densely that quantum mechanics influences their motions via the so-called degeneracy pressure.
When the compression is continued, the protons and neutrons will follow, providing a degeneracy pressure (such as in the core of neutron stars).
The equation of state at these densities is still unknown.
Beyond even the densities reachable in the interiors of neutron stars, we know that the final support of an ultra-dense matter is given by quarks via their degeneracy pressure (such conditions could perhaps exist inside the most massive neutron stars) \citep{annala_2020, annala_2021}.

\subsection{Degeneracy pressure}

Nuclear matter behaves in a somewhat analogous manner to liquids:
the particles composing the liquid experience attraction between their neighbors but strongly resist compression when tightly packed.
This resistance originates from zero-point motions of electrons (electron degeneracy pressure) and later on from zero-point motions and mutual interactions of neutrons (neutron degeneracy pressure).

The degeneracy pressure is based on Pauli's exclusion principle that forbids two identical fermions (spin-$\frac{1}{2}$ particles) from occupying the same quantum state.
This introduces a quantum mechanical pressure as fermions repel their neighbors with similar quantum states.
We can derive the resulting pressure from Heisenberg's uncertainty principle, which asserts that position and momentum can not be simultaneously determined to be better than
\begin{equation}
 \Delta x \Delta y \Delta z \Delta p_x \Delta p_y \Delta p_z = h^3,
\end{equation}
where $h$ is the Planck's constant, and $\Delta \vec{x}^3 \Delta \vec{p}^3$ is the 6D phase-space volume of the particles, where location vector $\vec{x} = (x, y, z)$, and momentum vector $\vec{p} = (p_x, p_y, p_z)$.
This means that there exists a zero-point momentum, called Fermi-momentum, $p_\mathrm{F}$.

The Fermi-momentum can be associated with a Fermi-energy, $E_\mathrm{F}$.
For non-relativistic case ($p \ll m c$) it is $E_\mathrm{F} = p_\mathrm{F}^2/2m$.
Here we consider Fermi energy for a (cold) 3D space with $N$ fermions filling the available energy states from the lowest up to the Fermi energy. 
It can be approximated similar to the energy in a square-well potential of size $L$,  
\begin{equation}
    E_\mathrm{F} = \frac{\pi^2 \hbar^2}{2 m L^2} (n_x^2 + n_y^2 + n_z^2) = \frac{\pi^2 \hbar^2}{2 m L^2} r_n^2
\end{equation}
where $r_n$ is the radius in $n$-space, and $n_i$ are the principle quantum numbers in 3D.
The total energy is obtained by integrating all states inside the Fermi sphere,
\begin{equation}
    E_\mathrm{tot} = 2 \frac{\pi^2 \hbar^2}{16 m L^2} \int_0^{r_n} 4\pi r_n^2 d r_n ~r_n^2 = \frac{\pi^3 \hbar^2}{10 m L^2} r_n^5
\end{equation}
Here the factor of $2$ comes from the fact that we have two different spin states for fermions.

\begin{figure}[t]
\centering
\includegraphics[clip, trim=0.0cm 0.0cm 0.0cm 0.0cm, width=0.45\columnwidth]{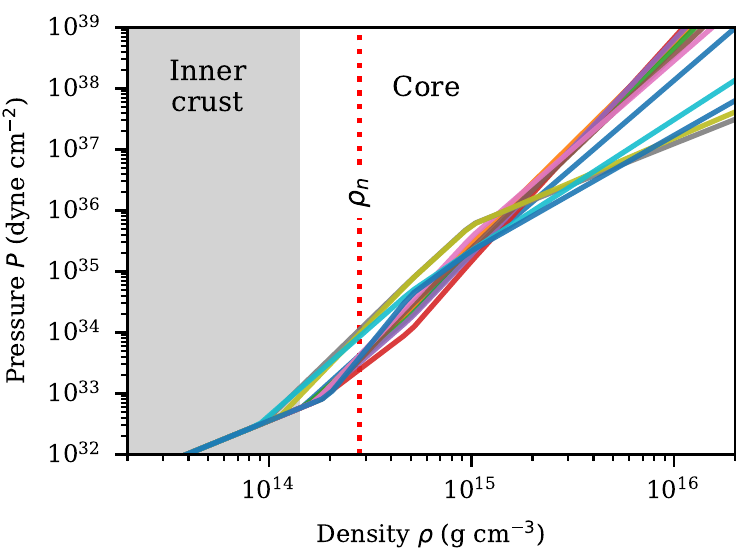}
\includegraphics[clip, trim=0.0cm 0.0cm 0.0cm 0.0cm, width=0.45\columnwidth]{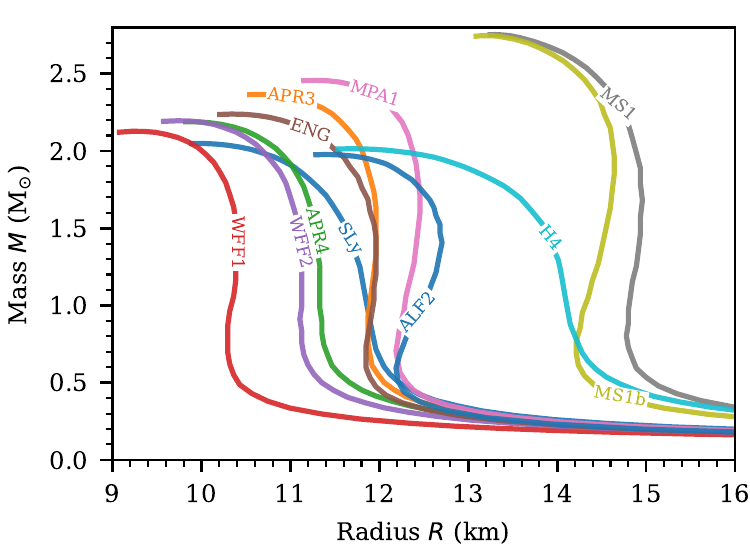}
\caption{\label{fig:eos_mr}
Examples of the different equation of state models and the corresponding neutron star mass-radius relations. 
Each EOS has a one-to-one mapping between the density-pressure and mass-radius planes.}
\end{figure}

Finally, the pressure is obtained from the standard thermodynamic relation (assuming a volume of $V = L^3$),  
\begin{equation}
    P = -\left( \frac{ \partial E_\mathrm{tot} }{\partial V} \right)_{S,N} 
    = \frac{\pi^3 \hbar^2}{15 m} \left( \frac{3N}{\pi V} \right)^{5/3}
\end{equation}
assuming constant entropy ($S$) and particle number ($N$).
We can generalize the expression into a form \citep{haensel_2007},
\begin{equation}
    P = P_r \frac{x_r^{3 \gamma_\mathrm{AD}} }{\gamma_\mathrm{AD}}
    \sim
    \begin{cases}
    P_r x_r^5 & (x_r \ll 1) \\
    P_r x_r^4 & (x_r \gg 1),
    \end{cases}
\end{equation}
where $P_r = m_e c^2/9 \pi^2 \bar{\lambda}_C^3 \sim 1.5 \times 10^{23} \mathrm{dyn} \cm^{-2}$ is a typical pressure, 
$\bar{\lambda}_C = \hbar/m_e c= 3.86 \times 10^{-11} \cm$ is the (reduced) Compton wavelength,
$x_r = p_\mathrm{F}/m_e c$ is the dimensionless Fermi-momentum,
and $\gamma_\mathrm{AD} = \frac{5}{3}$ or $\gamma_\mathrm{AD} = \frac{4}{3}$ is the polytropic index for non-relativistic ($x_r \ll 1$) or ultra-relativistic ($x_r \gg 1$) gas.
Note that $x_r \propto n_e^{1/3} \propto \rho^{1/3}$.
The transition between non-relativistic and ultra-relativistic regimes occurs at $x_r \sim 1$ corresponding to $\rho \sim 10^6 \gram \cm^{-3}$.

\begin{figure}[t]
\centering
\includegraphics[clip, trim=0.0cm 0.0cm 0.0cm 0.0cm, width=12cm]{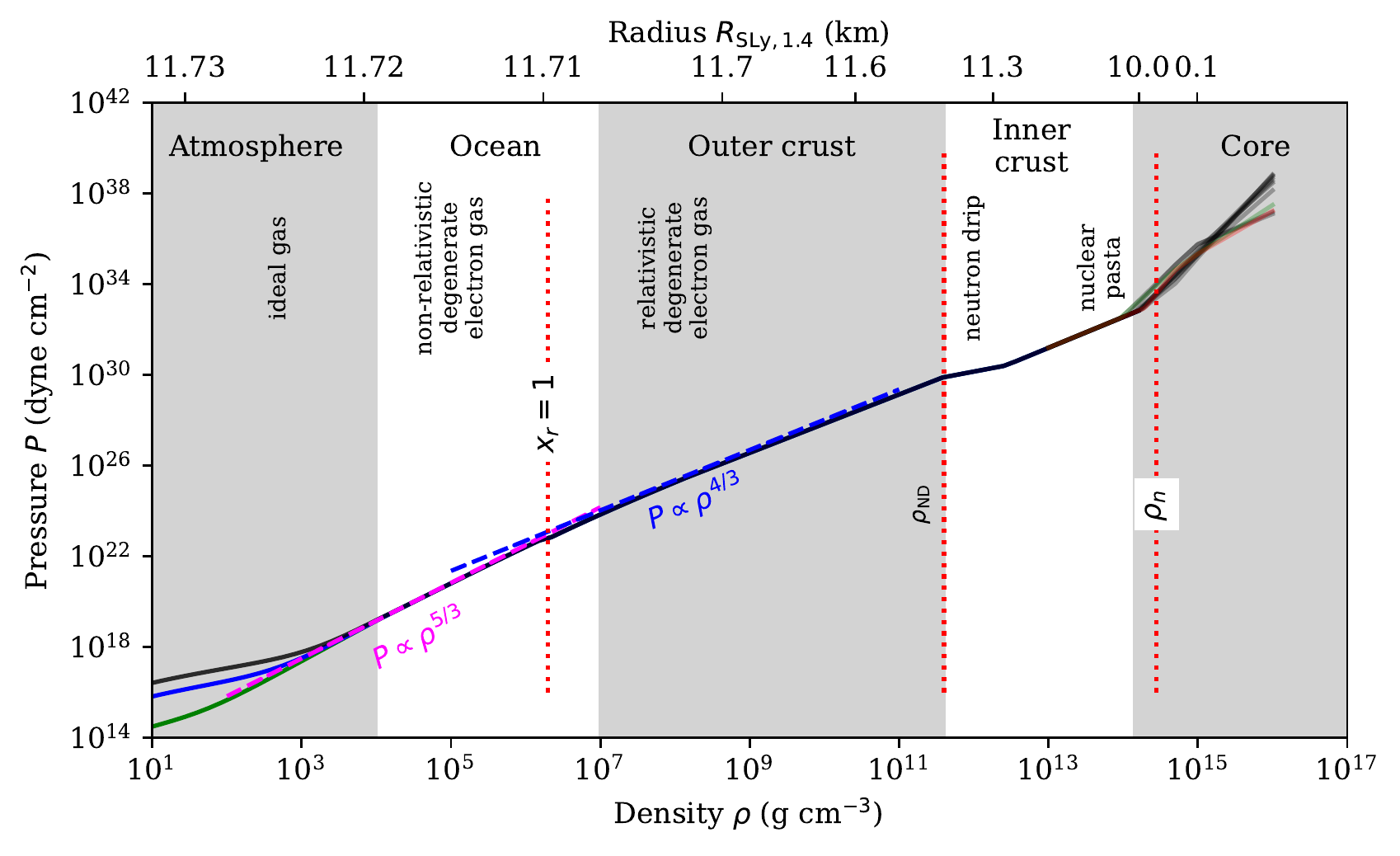}
\caption{\label{fig:eos}
    Equation of state, $P(\rho)$, of the dense matter inside neutron stars.
    The upper horizontal axis gives the same relation as a function of radius, $P(R_{SLy, 1.4})$, for a $M = 1.4\Msun$ star with the SLy equation of state. 
    Different regions inside the star are highlighted, including:
    atmosphere, for which the equation of state is given by the ideal gas law (various temperatures of $T = 10^6$, green curve; $5 \times 10^6$, blue curve; and $10^7 \Kelvin$, black curve are shown),
    ocean, where the plasma is Coulomb liquid, 
    outer crust, where the equation of state is split between non-relativistic degenerate electron gas ($P \propto \rho^{5/3}$; magenta dashed line) and relativistic degenerate electron gas ($P \propto \rho^{4/3}$; dashed blue line), and 
    inner crust and the core where the $P(\rho$) is still uncertain (various models are shown with different colored curves).
    Location of the neutron drip, $\rho_{\mathrm{ND}}$ (where neutrons start to leak out from the atomic cores) and saturation density, $\rho_\mathrm{n}$ (corresponding roughly to densities inside nucleons), are also shown with red dotted vertical lines.
  }
\end{figure}

This theoretical treatment provides a pathway to understanding the nuclear physics of the cores of neutron stars.
It is easy to see that degenerate neutron gas will behave similarly to degenerate electron gas since we can replace the mass $m_e \rightarrow m_n$.
Repulsive nuclear interaction, however, also contributes to the pressure at very high densities, rendering the equation of state in the neutron star core, $\rho > 10^{14} \gram \cm^{-3}$, unknown.
Multiple competing nuclear equation of state models exist at these densities (see Fig.~\ref{fig:eos_mr}).
Structure of a neutron star with one of these models, SLy (Skyrme Lyon model; \cite{2001A&A...380..151D}), is demonstrated in the previous Fig.~\ref{fig:eos}.

Most models assume fully nucleonic material (i.e., interactions between the neutrons).
Some models also assume that the material is enriched with hyperons, nucleons containing strange quarks or that the stars are composed entirely of strange-quark matter (mixture of up and down and strange quarks), making them, in fact, strange stars, in comparison to neutron stars.
The strange star models produce very soft equations of states which are generally disfavored by current observations of the heaviest neutron stars of $M \approx 2 M_{\odot}$ \citep[e.g.,][]{2010Natur.467.1081D}.

In summary, the state of ultra-dense matter is still open at the densities corresponding to the interiors of neutron stars.
Eventually, for high enough densities, the material will decompose into quarks, with an equation of state given by the perturbative quantum chromodynamics.
The equation of state in different density intervals is then composed of the following:
\begin{enumerate}[label=\roman*]
    \item ideal gas law, ($P = N kT/V$, where $kT$ is the gas temperature) for $\rho \lesssim 10^3 \gram \cm^{-3}$, 
    \item non-relativistic degenerate electron gas ($P \propto \rho^{5/3}$) up to $\rho \sim 10^6 \gram \cm^{-3}$,
    \item relativistic degenerate electron gas ($P \propto \rho^{4/3}$) up to $\rho \sim 10^{11} \gram \cm^{-3}$,
    \item degenerate neutron gas with additional, unknown, nuclear interactions, and, eventually,
    \item quark matter with pressure given by quantum chromodynamics, at $\rho \gtrsim 40 \rho_n$ (where $\rho_n$ is the saturation density).
\end{enumerate}
The equation of state is also visualized in Fig.~\ref{fig:eos}.
Especially the unknown equation of the state of the matter inside the core is under active research.

Observations of neutron stars allow an unrivaled opportunity to put definite constraints on the state of the ultra-dense matter.
Various methods exist for constraining the mass, both mass and radius, and in some cases, the state of the matter indirectly.
For recent reviews, see \citep{ozel_2016rev, suleimanov_2016,  miller_2016, lattimer_2019} and especially \cite{degenaar_2018} that focuses specifically on the various methods for mass and radius measurements.
We have already discussed the possibility of using the pulse profiles (and their dependency on the gravitational potential) to provide constraints for the equation of state.
Other promising methods include constraining the size of the emission directly from the X-ray emission emanating from the surface.
Various methods used to constrain the equation of state of ultra-dense matter inside neutron stars are listed and described in Table~\ref{tab:measurements}.

\subsection{Thermal-like emission from isolated neutron stars}

The most important part of interpreting the electromagnetic emission from neutron stars is understanding the physics of the atmosphere that re-processes all the radiation going through it.
The atmosphere of a neutron star is a thin, $\sim 1$ to $10\cm$ height, a slab of hot plasma.
To model it, equations of radiative transfer are solved, and the atmosphere is (typically) assumed to be in a radiative and hydrostatic equilibrium.
This kind of modeling bears a great similarity to modeling regular stellar atmospheres.

The atmosphere is thought to be composed of light elements. 
It is typically assumed to be in a local thermodynamical equilibrium (LTE) so that the equation of state follows an ideal gas law, $P = n k T$ and Saha equation can be used to calculate the ionization levels.
Heavier elements are expected to sink to the bottom due to the large gravitational potential (i.e., the atmosphere has a very short sedimentation timescale);
hence the atmosphere is often taken to be composed of the lightest elements only, such as hydrogen or helium.
For accreting systems, the composition is often taken to coincide with the elements in the companion star's outer envelope.

The most direct application of the atmosphere models is the interpretation of the thermal-like emission from central compact objects (CCOs),  point-like X-ray sources in supernova remnants, using the atmosphere model spectra.
These CCOs are young ($t < 10^4\yr$) neutron stars with modest X-ray luminosities of $L_x \sim 10^{33}\ergs$.
They have thermal spectra consistent with temperatures of $kT \sim 0.1$ to $0.5\kev$.
Interestingly, recent observations indicate that some CCOs are better explained by the carbon composition of the atmosphere.
Using carbon atmosphere models, Ho \& Heineken \cite{2009Natur.462...71H} constrained the radius of a neutron star in Cas A to be between $R \approx 10 - 14\km$.
Similar measurement was done for HESS J1731-347 by Klochkov et al. \cite{2014ApJS..210...13S, 2015AA...573A..53K} who constrained the mass and radius to be $M = 1.55\pm 0.3\Msun$ and $R=12.4^{+0.9}_{-2.2}\km$, respectively.

\newgeometry{margin=1cm} 
\begin{landscape}
\begin{table}[ht]
\caption{Methods for constraining the neutron star equation of state}
\begin{center}
 \begin{tabular}{ p{0.15\linewidth} | p{0.45\linewidth}| p{0.25\linewidth}}
    \textbf{Method} & \textbf{Description} & \textbf{Notable measurements}\\[3ex]
\multicolumn{3}{c}{\textit{Mass measurements}}\\[3ex]
Radio pulsars in binary systems & Accurate timing of radio pulsars allow the determination of orbital parameters (orbital period $P_{\mathrm{orb}}$, inclination angle $\sin i$, the mass of the companion $M_c$, and total mass of binary $M_T = M_{NS} + M_c$) for neutron stars in binary systems. 
For nearly edge-on systems, the propagation of the signal through the gravitational potential of the companion allows breaking the degeneracy in the mass (Shapiro delay). & 
PSR~J161-2230: $M = 1.908 \pm 0.016 \Msun$ \cite{2010Natur.467.1081D, 2016ApJ...832..167F,2018ApJS..235...37A},
PSR~J0348+0432: $M = 2.01 \pm 0.03 \Msun$ \cite{2013Sci...340..448A},
PSR~J0740+6620: $M = 2.08 \pm 0.07 \Msun$ \cite{2021ApJ...915L..12F}.
 \\[2ex]
\hline
  Optical measurements of binary systems &  Optical measurements of a companion star rotating around a neutron star allow the determination of orbital parameters ($P_{\mathrm{orb}}$, $\sin i$, $M_c$, and $M_T = M_{NS} + M_c$) of the system. Additional information via \textit{i}) Shapiro delay, \textit{ii}) observations of spectral lines of the companion star, or \textit{iii}) modeling of eclipses (black widow systems) can be used to break the degeneracy between the mass and inclination. & Black widow system PSR~J0952-0607 $M = 2.35 \pm 0.17 \Msun$ \cite{2022ApJ...934L..18R}. \\[8ex]
\multicolumn{3}{c}{\textit{Simultaneous mass and radius measurements}}\\[3ex]
   Thermonuclear X-ray bursts & Time-resolved X-ray observations of cooling of neutron star atmosphere after thermonuclear X-ray bursts allow determination of $R$, $M$, distance $D$, hydrogen mass fraction $X$, and other system parameters by comparing the cooling track to theoretical model calculations. 
   Distance measurements help narrow the constraints. & Burster 4U~1702$-$429 $R = 12.4\pm 0.4\km$ and $M = 1.9\pm 0.3\Msun$ \cite{nattila2017}. \\[3ex]
      \hline
   X-ray pulse profile modeling & Modeling of phase-folded X-ray pulse profiles allow constraining the parameters of the gravitational potential (compactness $GM/Rc^2$). Higher-order effects like Doppler boosting and relativistic aberration of angles allow breaking of the degeneracy between $R$, $M$, and observer inclination angle $\sin i$ and spot co-latitude $\cos\theta_s$. & PSR~J0030$+$0451 $R \approx 12.7\pm 1.2\km$ and $M \approx 1.34 \pm 0.16 \Msun$ \cite{2019ApJ...887L..21R} or
$R \approx 13.0^{+1.3}_{-1.1} \km$ and $M\approx 1.44 \pm 0.15 \Msun$ \cite{2019ApJ...887L..24M}.
PSR~J0740$+$6620 with mass $M \approx 2.08 \pm 0.07 \Msun$ and $R \approx 12.4^{+1.3}_{-1.0} \km$ \cite{2021ApJ...918L..27R} or $R \approx 13.7^{+2.6}_{-1.5} \km$ \cite{2021ApJ...918L..28M}. \\[3ex]
   \hline
   Thermally-emitting qLMXBs & Modeling of X-ray spectral observations of quiescent (non-accreting) X-ray binary systems allow constraining the size of the emitting region ($\propto R^2/D^2$) and relativistic effects carry additional information about compactness which allows weak constraints on the mass $M$ too. 
   Distance measurements help narrow the constraints. & 
   Statistical combination of five sources yielded $R = 9.1_{-1.5}^{+1.3}\km$ for $M = 1.4\Msun$ (90\% confidence) \cite{2013ApJ...772....7G}.
Similar analysis of 8 sources yielded $R = 10 - 14\km$ for $M = 1.4\Msun$ \cite{2018MNRAS.476..421S}. \\[3ex]
   \hline
   Thermally-emitting isolated neutron stars & Modeling of X-ray spectra observations of isolated neutron stars allow constraining the size of the emitting region ($\propto R^2/D^2$), and relativistic effects carry additional information about the compactness, which allows weak constraints on the mass $M$ too. Distance measurements help narrow the constraints.  & CCO HESS~J1731$-$347 $R = 12.4^{+0.9}_{-2.2} \km$ and $M = 1.55\pm 0.3\Msun$ \cite{2015AA...573A..53K}.
  \\[3ex]
\multicolumn{3}{c}{\textit{Other constraints}}\\[3ex]
   Constraints from fast spin & Measurements of neutron stars with fast spin allow constraining the maximum possible values of $M$ and $R$ by requiring the star not to break up (mass shedding limit.) & PSR~J1748$–$2446a rotating $f = 716.356\Hz$ \cite{2006Sci...311.1901H} limits $R < 16\km$ (for $M = 2\Msun$). \\[3ex]
   \hline
   Tidal deformability measurements & Modeling of gravitational wave signals from mergers carry information on the tidal deformability of the material that can be translated to $M$ and $R$ constraints. & GW170817 allowed constraining the tidal deformability \cite{2017PhRvL.119p1101A}, which in turn translates to $R \lesssim 14\km$ for $M = 1.4\Msun$. \\[3ex]
   \hline
\end{tabular}
\end{center}
\label{tab:measurements}
\begin{center}
\end{center}
\end{table}
\end{landscape}
\restoregeometry

\subsection{Thermonuclear X-ray bursts}

Thermonuclear (type-I) X-ray bursts are a phenomenon unique to neutron stars (for a recent review, see \cite{2021ASSL..461..209G}). 
They offer a more complex alternative for measuring the neutron star radius by comparing the time-resolved X-ray spectra to theoretical atmosphere models.
In addition, thermonuclear burning on top of neutron stars is a rich phenomenon with many unknown physical features.

The bursts are triggered by unstable thermonuclear burning of matter accumulated onto the neutron star envelope from a binary companion.
During the burning, hydrogen and helium fuse into heavier elements through the hot CNO cycle and triple alpha reactions, sometimes triggering the rp-process (rapid proton captures) that can generate elements up to Tellurium \cite{2001PhRvL..86.3471S}.
Enormous amounts of energy are released quickly, particularly for helium-rich fuel: triple alpha reactions are not limited by beta decays that limit the CNO hydrogen burning.
The most energetic bursts can have rise times that are less than an ms \cite{2014A&A...568A..69I}, suggesting that on rare occasions, the neutron star envelope can also detonate rather than deflagrate (i.e., the burning front can expand both super- and sub-sonically), and cause a relativistic outflow that drives the outer neutron star envelope into the interstellar space.

Observations of X-ray bursts probe several key fundamental physics problems.
In bursts where the rp-process is important, the (often poorly known) nuclear reaction rates of various proton-rich isotopes play a key role in the morphology of the X-ray light curves \cite{2016ApJ...830...55C}. 
This has the potential to make the thermonuclear bursts also laboratories of nuclear physics, rivaling any Earthly nuclear burning physics experiment on nuclear reaction rates.
The extreme parameter regime also enables studies of more exotic proton-rich isotopes that can be present during the burning.

The burning ashes generated during the X-ray burst --- in addition to the possible, stable nuclear burning \cite{1999ApJ...524.1014S} --- settle down to the neutron star ocean. 
At the bottom of the ocean, chemical separation occurs, wherein heavier elements crystallize into a solid crust, and lighter elements such as oxygen and carbon preferably remain in the liquid ocean \cite{2007PhRvE..75f6101H}.
This transitional layer of the star is where so-called super-bursts are triggered, which are powered by nuclear burning of carbon \cite{2000A&A...357L..21C,2001ApJ...559L.127C,2002ApJ...566.1045S}.
In the accreted crust \cite{2018ApJ...859...62L}, the chemical composition is further altered through pycnonuclear reactions that also heat the crust \cite{2008A&A...480..459H}. 
However, the neutron-rich nuclei in the outer crust undergo repeated cycles of electron captures and beta decay. The resulting neutrino losses effectively thermally decouple the ocean/atmosphere from the hotter inner crust \cite{2014Natur.505...62S}.
Also, very recently, a new type of nuclear burning regime has been proposed, the so-called ``hyper bursts'', that are triggered possibly by unstable burning of oxygen or neon at densities of $\rho \sim 10^{11} \gram \cm^{-3}$ near the neutron drip \cite{2022arXiv220203962P}.
It is practically impossible to obtain direct observational constraints deeper down from these densities. Therefore, to probe the properties of the inner crust and the ultra-dense neutron star core, we must place constraints on the macroscopic properties of the star: their masses and the radii.

We can place constraints on the neutron star masses and radii by analyzing quasi-thermal light emitted by their photospheres. 
During X-ray bursts, the entire atmosphere heats up, such that the photospheric temperatures can reach almost 3 keV, and then it rapidly cools down as the fusion reactions start waning down at the burning depths.
The most energetic bursts reach the Eddington limit, which results in photospheric expansion, and they act as standard distance candles \cite{2003A&A...399..663K}.
Moreover, these photospheric radius expansion bursts are expected to launch radiatively driven winds \cite{1986ApJ...302..519P, 2021ApJ...914...49G}, which may enrich the interstellar medium with the burning ashes \cite{2006ApJ...639.1018W,2010A&A...520A..81I,2017MNRAS.464L...6K}.

Multiple techniques have been used to constrain neutron star masses and radii. The majority of them --- such as the touchdown method \cite{1990A&A...237..103D,2006Natur.441.1115O} or the cooling tail methods \cite{2011ApJ...742..122S,2017MNRAS.466..906S, nattila_2016} --- rely on the fact that their X-ray spectra can be adequately modeled with a simple black body model.
Another approach is modeling X-ray bursts using sophisticated atmosphere models, whose properties depend on the neutron star mass and radius, among many other variables \cite{nattila2017}.

All of these techniques, however, have relied upon the assumption that the environment where the X-ray bursts occur remains unaltered by the explosion of light and mechanical energy exerted by the burst-driven winds.
Recent observational data indicate clearly that this is not the case the majority of times; the X-ray bursts alter the properties of the accretion discs surrounding the neutron stars \cite{2013ApJ...772...94W,2015ApJ...801...60W,2016MNRAS.455.2004Z,2017A&A...599A..89K,2018SSRv..214...15D}. The accretion flow in the disc conversely seem to influence the spectral properties of the bursts \cite{2014MNRAS.445.4218K, 2017MNRAS.472...78K}.
The burst-disc interactions severely limit the use of X-ray bursts in making precise mass-radius measurements. Typically, their impact can only be inferred from studying large numbers of bursts rather than seeing clear imprints in individual bursts.
In addition, the atmosphere models that have been used so far ignore many important factors, such as rapid stellar rotation that can cause the star to become oblate and modify the energy distribution of the radiation via strong Doppler boosting.

But the potential of constraining dense matter equations of state models using X-ray bursts is enormous.
For example, \cite{nattila2017} constrained the neutron star's radius in one X-ray burster to within $400$ meters.
This target was, however, ideal in many ways: the mass accretion rate was uncharacteristically low, and the source was in the hard spectral state, facilitating optimal measurement conditions for the radius.
Unfortunately, most of the other observed bursts/bursters are less optimal in their behavior.

\newpage
\section{Laboratories of Electrodynamics}

Understanding the physics of neutron star magnetospheres requires the usage of both classical electromagnetism and quantum electrodynamics (QED)
\citep{arons_1979, arons_2007, cerutti_2017, beskin_2018}.
From the perspective of electromagnetism, the neutron star is just a rotating magnetized ball in a vacuum.
This kind of simplification, however, only gets us so far because the strong magnetic field also means that electromagnetic radiation (photons) and plasma particles start to a couple in the framework of QED.

\subsection{Spindown power of magnetized balls}

What can we expect from a dense ball of plasma ($R \sim 10^6\cm$) that is rotating rapidly ($P \sim 1\second$) and spinning a huge magnetic field ($B\sim 10^{12}\gauss$) around?
The rotational energy of such a star is given by 
\begin{equation}
E = \frac{1}{2} I \Omega^2,
\end{equation}
where the moment of inertia $I = \frac{2}{5} M R^2 \sim 10^{45} \gram \cm^2$.
Since isolated pulsars are observed to slow down, the corresponding spin-down power is
\begin{equation}
\dot{E} = I \Omega \dot{\Omega} \sim 10^{32} \ergs.
\end{equation}
for an angular velocity of $\Omega = 2\pi/P \approx 6 \second^{-1}$, 
and spin-down rate of $\dot{\Omega} \sim 10^{-13}$.

The simplest physical model for this slow-down is an electromagnetic torque acting on a rotating dipole. 
The angular momentum is carried away from the system by electromagnetic radiation, Poynting flux. 
The corresponding magnetic dipole energy loss rate is
\begin{equation}
    \dot{E} = \frac{2}{3}\frac{ (\ddot{m}_\perp)^2}{c^3} = \frac{2}{3} \frac{B^2 \Omega^4 R^6}{c^3} \sin^2\chi \sim 10^{32} B_{12}^2 P_0^{-4} \ergs,
\end{equation}
where $m_\perp = m \sin\chi$, 
$m = B R^3$ is the magnetic moment,
$B$ the star's polar magnetic field, 
and $\chi$ is the magnetic dipole inclination angle with respect to the spin axis.
The period and angular velocity are related simply as $P = 2\pi/\Omega$.
In the last expression, the quantities are scaled as $Q_x \equiv 10^x Q$.

Equating the spin-down power and the magnetic dipole losses allows us to estimate the magnetic field.
We obtain,
\begin{equation}
    B \approx 10^{12} \sqrt{\frac{P}{1\second}} \sqrt{\frac{\dot{P}}{10^{-15}}} \gauss,
\end{equation}
that agrees with our previous estimate of the strength of the magnetic field for characteristic pulsar parameters.

\subsection{Charges in the magnetosphere}

The above estimates demonstrate that the basic arguments based on electrodynamics lead to quantitative agreement.
The standard pulsar model, therefore, consists of a rotating magnetized ball with a dipole field. 
Let us now refine this model to include charge carriers --- plasma.

From the perspective of our ``rotating ball model'', the magnetic field is spun by an excellent conductor.
This rotation induces electric fields in the conductor.
We can model the pulsar as a Faraday disk dynamo to estimate the order of magnitude of these electric fields.
In this picture, we have a disk of radius $R$, rotating with an angular velocity $\Omega$.
The disk is penetrated by magnetic fields of strength $B \sim 10^{12} \gauss$.
The rotating conductor experiences the moving magnetic field as an electric field, 
\begin{equation}
    |\vec{E}| = |-\frac{\vec{v}}{c} \times \vec{B}| \sim \frac{\Omega B R}{c}, 
\end{equation}
where $\vec{v} = R \nvec{r} \times \Omega$ is the tangential velocity of the outer disk rim.
The corresponding voltage induced by the rotating disk (for a Faraday disk generator) is $V \sim 10^{15} \mathrm{V}$.
There is also a large current flowing in the ``circuit'', $I = P/V \sim 10^{14} ~\mathrm{A}$ (assuming spin-down power of $P \sim 10^{38} \ergs$).

The large electric field can lift charges from the star's surface.
Goldreich \& Julian \cite{goldreich_1969} showed that this induces a (minimum) charge density of 
\begin{equation}
\rho_\mathrm{GJ} = \frac{\nabla \cdot \vec{E}}{4\pi} \approx -\frac{\Omega \cdot \vec{B}}{2\pi c}.
\end{equation}
These charges are pulled from the surface because the electromagnetic forces far exceed gravity.
The corresponding number density of plasma is $n_\mathrm{GJ} = \rho_\mathrm{GJ}/e \sim 10^{12} \cm^{-3}$.
An actual number density can be even higher, $n = \mathcal{M} n_\mathrm{GJ}$, where $\mathcal{M} \gg 1$ is the so-called multiplicity parameter.

The charge density in the magnetosphere leads to a formation of a (minimum) conduction current of 
\begin{equation}
    \vec{J} = n e \vec{v},
\end{equation}
where $n e = \rho_R$ is the charge density required to screen the parallel electric 
\begin{equation}
    E_\parallel = \frac{\vec{E} \cdot \vec{B}}{B} \rightarrow 0;
\end{equation}
in the opposite case, the electric field would lift the charges until it would become screened.

The minimum Goldreich-Julian charge density \citep{goldreich_1969} also has higher order relativistic corrections so that $\rho_\mathrm{GJ} \rightarrow \rho_\mathrm{GJ} + $ additional relativistic terms \citep{muslimov_1992}.
They modify the Goldreich-Julian charge density, 
\begin{equation}
    \rho_\mathrm{GJ} \approx -\frac{(\Omega -\omega) B}{2\pi c},
\end{equation}
where $\omega \approx \frac{2 G I}{c^2 r^3} \Omega$ is the angular velocity of the spacetime drag at a distance $r$ from a rotating body.
These corrections are, in reality, important to fully explain the plasma feeding into the magnetosphere.

\subsection{Force-free and magnetohydrodynamic solutions}

\begin{figure}[t]
\centering
\includegraphics[clip, trim=0.0cm 0.0cm 0.0cm 0.0cm, width=12cm]{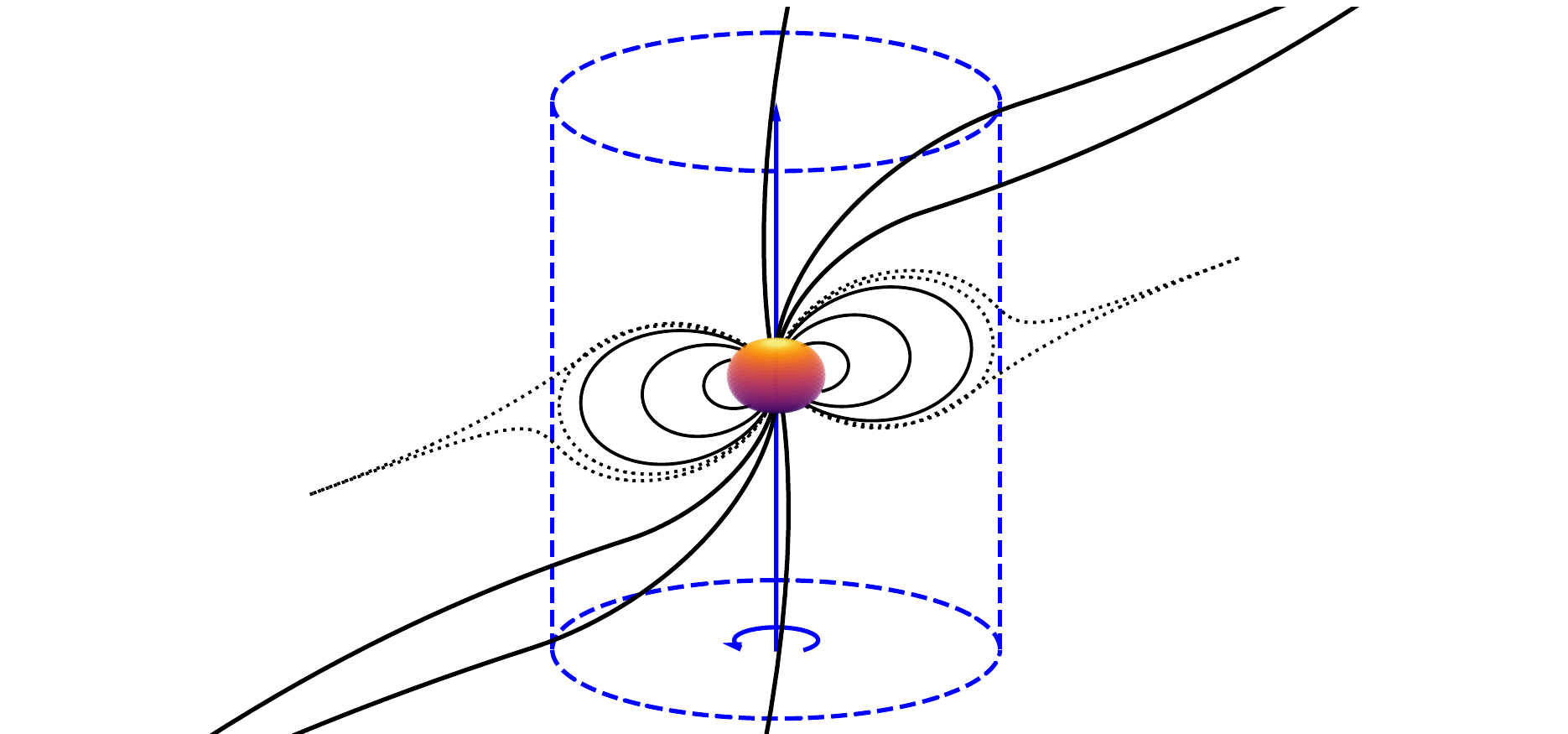}
\caption{\label{fig:pulsar}
    Schematic view of the pulsar magnetosphere.
    The neutron star in the center has an inclined dipole magnetic field w.r.t. to the spin axis of $\chi = \pi/12$.
    Open and closed magnetic field lines are shown with solid black curves;
    magnetic field lines forming the Y-point are depicted with dotted curves.
    The light cylinder is shown with blue dashed curves.
  }
\end{figure}

Much of the further progress in understanding the structure of the magnetospheres originates from numerical studies.
This is understandable because the related equations are nonlinear and often require numerical tools to probe the solutions.
The subsequent structure of the magnetosphere depends on how dense the plasma is, that is, if $\mathcal{M} \gg 1$.

Goldreich \& Julian \cite{goldreich_1969} introduced the idea of a force-free magnetosphere where the parallel electric fields are always screened.
Here, the electromagnetic energy density is so large that all inertial, pressure, and dissipative forces can be neglected.
This means that the pulsar can be considered immersed in a massless conducting fluid instead of a vacuum.
Similarly, Michel \cite{michel_1969} introduced a magnetohydrodynamical counterpart; 
this also assumes a large density, compared to $n_\mathrm{GJ}$.

In both systems, the plasma starts to rotate rigidly together with the star.
This rigid rotation becomes impossible at a sufficiently large radius.
This is known as the light cylinder, where the magnetic field moves with the speed of light, and the plasma can no longer co-rotate.
Therefore, the light cylinder radius is
\begin{equation}
R_\mathrm{LC} = \frac{c}{\Omega}.
\end{equation}
It is $R_\mathrm{LC} \sim 10^{9}$ to $10^{10}\cm$ (i.e. $10^3 R_\star$) for $P \sim 0.1 - 1\second$.
Inside this region, $r < R_{LC}$, the magnetic field is primarily dipolar (higher order moments decrease faster with $r$ than dipole moment).
Outside of it, $r > R_{LC}$, all the field lines are open, and the flux is conserved so that $B r^2 = \mathrm{const.}$;
then, $B \propto r^{-1}$.

The problem of magnetospheric structure can be formulated in a type of Grad-Shafranov equation that encapsulates the aligned rotator;
the equation is familiar from magnetic confinement \citep{bateman_1978} with Grad-Shafranov equation being its relativistic counterpart \citep{scharlemann_73, michel_1973}.
Contopolous et al. \cite{contopoulos_1999} solved this equation by iterative technique assuming $\vec{E}\cdot\vec{B} = 0$ and $E^2 - B^2 < 0$ (implying subluminal $E\times B$ drift velocity).
The important outcome of this exercise was that the last closed field at the equator has a Y-type neutral point.
This means that the return current flows in a current sheet at the boundary of the closed zone where the magnetic field flips sign.
The result has since then been produced by others confirming its universal nature \citep{gruzinov_2005, mckinney_2006, timokhin_2006, spitkovsky_2006}
The corresponding spindown energy losses of the aligned rotator were numerically confirmed to be 
\begin{equation}
    \dot{E} = k \frac{ \Omega^4 \mu^2}{c^3},
\end{equation}
with $k = 1 \pm 0.1$.

\subsection{Evolving magnetic topology}

Important progress came when Spitkovsky \cite{spitkovsky_2006} numerically solved the magnetospheric structure with time-dependent force-free simulations.
This confirmed the presence of the current sheet at the equator and allowed the solution with an arbitrary (magnetic) inclination angle $\chi > 0$ (angle between the magnetic moment and rotation axis).
Then, the generalized spindown is
\begin{equation}
    \dot{E} = k \frac{\mu^2 \Omega^4 }{ c^3 } (1 + \sin^2 \chi)
\end{equation}
with $k = 1 \pm 0.1$.
This expression introduces the $\dot{E} \propto (1 + \sin^2\chi)$ dependency on the inclination angle, $\chi$.
The magnetic topology of an oblique rotator, $\chi > 0$, is just a rotationally distorted version of the simpler aligned rotator geometry, $\chi = 0$.
Note that the solution still assumes a magnetosphere filled with the plasma of density sufficient to short out the parallel electric fields.

This generalized spindown power formula suggests an increase in the losses with increasing $\chi$.
Therefore, the inclination angle should decrease with time for old pulsars.
Since $R_\mathrm{LC}$ increases as the pulsar spin downs, there must be a net conversion of open field lines to closed field lines on the spin-down timescale.
This topological change requires the violation of the ideal MHD.
This will then necessarily lead to reconnection \citep{lyubarsky_2001}.
The most likely location is at the Y-point and in the current sheet.
The importance of reconnection was later confirmed with fully kinetic simulations that do not require us to impose the assumption of dense plasma background
\citep{philippov_2014a, philippov_2014b, chen_2014, philippov_2015, philippov_2015b, cerutti_2015, cerutti_2016}.

\subsection{Mysterious Pulsar Radio Emission}

The spin-down luminosity, $\sim 10^{38} \ergs$, leaks out from the pulsar magnetosphere and can act as free energy to produce observable emission.
However, only a tiny fraction, $10^{-6}$ to $10^{-4}$ of that energy is known to end up in the radio band to generate the actual pulsar emission \cite[see][for a recent review]{philippov2022}.
Quoting J. Arons, this makes pulsars ``\textit{dogs that don't bark}'' \cite{arons_1979}.
The origin of this emission is still unknown, but it is thought to be connected to the pair avalanches (see next section) regulating the electric circuit of the magnetosphere.

The radio emission is broadband, extending to as low as tens of MHz and up to 150 GHz.
The energy output peaks at a frequency of a few hundred MHz with a specific energy flux of a few Jansky \cite{philippov2022}.
\footnote{
Jansky is a unit of energy flux used primarily in radio astronomy.
It is $10^{-23}\ergs\,\cm^{-2}\,\mathrm{Hz}^{-1}$.}
The emission is elliptically polarized with a high degree of polarization, reaching $\approx 100\%$.
The linearly polarized component dominates the emission.
Circular polarization degree can reach up to $\approx 10\%$.
The polarization angle of many pulsars has a canonical swing, where the angle evolves as a function of the pulsar phase, tracing a (tilted) S-shaped trajectory.
This angle sweep takes place independently of frequency, pointing to geometric effects that are well explained with a rotating vector model \cite{1969ApL.....3..225R, 2020A&A...641A.166P}.

Observations averaging over multiple pulses of radio emission have given the pulsars their status as the most accurate clocks in the Universe.
Individual pulses, however, vary significantly from pulse to pulse, and many show high sub-pulse variability.
Some pulsars have individual pulses that drift in phase, others sometimes switch off (nulling), and others abruptly flip between various states (moding).
The origin and nature of these sub-pulse variabilities are still unknown.

\subsection{Pulsar wind nebulae}

So where does the rest of the large spin-down energy go?
Most of that energy is carried away by electromagnetic Poynting flux and by the charged-particle outflow streaming out from the pulsar.
But all that energy and charged particles are not just freely flowing into a vacuum;
the neutron stars are surrounded by gas from their old stellar remnants.
The interaction of that electromagnetic energy and charged particles generates a hazy, glowing turbulent blob of gas around the neutron star known as a pulsar wind nebula.
This makes the nebulae huge calorimeters in the sky that trap, reprocess, and weigh the energy output of pulsars.

As an example, the Crab nebula is probably the most well-studied nebula (and neutron star) of all time.
Crab has a spin period of $P \approx 33 \,\mathrm{ms}$ and a period derivative of $\dot{P} \approx 420 \times 10^{-15}$.
Its age can be estimated from $P/2\dot{P} \approx 1000 \yr$ (dynamical age), matching well with the historical supernova in 1054AD.
The nebula emits in a broad range of spectrum from radio to optical to X-rays to gamma-rays;
injected powered of $\dot{E} = 5 \times 10^{38} \ergs$ is required to be channeled into the surrounding plasma to explain the observed luminosity of the nebula.
These estimates are well in-line with the spin-down power of a magnetized ball with $B \sim 10^{12}\gauss$, corresponding to a power of $\sim 10^{38} \ergs$.

To better understand the interaction of energy and the charged particles, the radiation and its generation, we need an understanding also of plasma physics. 
This is the focus of our next section.

\newpage
\section{Laboratories of plasma physics}

The magnetospheres of neutron stars are composed of charged particles --- plasma.
As the magnetospheric plasma cools down, it radiates its energy away and populates the medium with photons.
The nonlinear dynamics of this plasma and radiation are governed by radiative plasma physics.
In this section, we introduce the basic concepts of such systems.

\subsection{Standard quantum electrodynamic interactions} 

The magnetospheres of regular radio pulsars are composed of a large magnetic field of $B_\star \sim 10^{12}\gauss$.
Rapid rotation of this field leads to a generation of large electric fields, as well;
these electric fields can open large potential gaps in the magnetosphere.
The resulting voltage accelerates particles to relativistic energies, and these high-energy particles produce luminous radiation.
The interaction of the charged particles and the radiation field requires the coupling of radiative physics and electrodynamics.
Therefore, the pulsar magnetospheres are prime laboratories of quantum electrodynamics \citep{jauch_1976, berestetskii_1980}.

In practice, quantum electrodynamics describes the interaction of photons (electromagnetic radiation) and charged particles.
The simplest two-body interactions between photons and electrons are:%
\footnote{We give a schematic description of these interactions by denoting electrons with $e^-$, positrons with $e^+$, and photons with $\gamma$.
    Pairs are succinctly given as $e^\pm$.
    Altered (or often energized) end-products are denoted with a prime, $x \rightarrow x'$.
    The passive magnetic field (or virtual photons) is expressed with $[B]$.
}
\begin{enumerate}[label=\roman*]
    \item  pair creation (photons interacting with photons to create pairs), $\gamma + \gamma' \rightarrow e^- + e^+$;
    \item pair annihilation (electrons interacting with positrons to create photons), $e^- + e^+ \rightarrow \gamma + \gamma'$; 
    \item Compton scattering (photons interacting with pairs to exchange energy), $e^\pm + \gamma \rightarrow {e^\pm}' + \gamma'$; and
    \item Coulomb scattering (pairs interacting with pairs to exchange energy), $e^\pm + e^\pm \rightarrow e^{\pm \prime} + e^{\pm\prime\prime}$ and $e^\pm + e^\mp \rightarrow e^{\pm\prime} + e^{\mp\prime\prime}$.
\end{enumerate}
Magnetic field adds an extra flavor to these (lowest-order) interactions because it can be thought of as consisting of virtual photons.
This enables single-body interactions (or, alternatively, two-body interactions composed of ``real'' against ``virtual'' species) such as
\begin{enumerate}[label=\roman*,resume]
    \item synchrotron radiation (pairs upscattering virtual photons), $e^\pm + [B] \rightarrow e^\pm + \gamma + [B]$; and
    \item single-photon (magnetic) pair creation (photons pair-creating with virtual photons), $\gamma + [B] \rightarrow e^\pm + [B]$;
\end{enumerate}
All of these processes are important in the neutron star magnetospheres.

\subsection{Pair cascades}

The magnetosphere of pulsars is prone to avalanches of pair creation \citep{sturrock_1971}.
They evolve as follows \citep{beskin_2018}.
The twisting electric field (from the rotation of the star) induces a longitudinal electric field to the magnetosphere, $E_\parallel \sim (\Omega R /c) B$.
Any (primary) particle in this region is accelerated to ultra-relativistic energies, $\varepsilon_e \gg 1$, where $\varepsilon_e \equiv E_e/m_e c^2$ is the particle's energy, $E_e$, in the units of electron-rest mass $m_e c^2$.

The strong magnetic field of the environment, however, renders the synchrotron cooling time to be very short,
\begin{equation}
    \tau_s \sim \frac{1}{\omega_B} \frac{c}{\omega_B r_e} \sim 10^{-15} \second,
\end{equation}
where $\omega_B = eB/m_e c$ is the particle's gyro-frequency and $r_e = e^2/m_e c^2$ classical electron radius.
Since the synchrotron emission power depends on the perpendicular component of the particle's velocity, $P_{synch} \propto \varepsilon_e^2 \sin\alpha$, where $\sin\alpha$ 
is the angle between the particle's momenta and the magnetic field.
This forces the particles to move along the magnetic field lines; pairs in the magnetosphere can be imagined as beads sliding on a wire.

The synchrotron cooling and the curvature of the magnetic field causes these electrons (or positrons) to emit gamma-ray photons, $\varepsilon_{\gamma} \sim 1$, as they slide along the $B$ field lines. 
Here $\varepsilon_\gamma \equiv E_\gamma/m_e c^2$ is the energy of the photon in the units of electron rest mass. 
The energy losses from the curvature radiation, $P_c = \frac{2}{3} \frac{e^2}{R_c^2} \varepsilon_e^4$, depend on the curvature radius, $R_c \sim R_{NS}$, and the energy of the electron, $\varepsilon_e$, and impose an upper bound on the particle's energy.
Balancing acceleration, $e E_\parallel$, and the curvature losses, $P_c$, results in $\varepsilon_\mathrm{max} \sim (R_c^2 E_\parallel/e)^{1/4}$ (i.e., $e_\mathrm{max} m_e c^2 \sim 10^7 \mathrm{MeV}$).
Therefore, the magnetosphere is expected to be full of high-energy gamma rays.

The gamma-rays, corresponding to energies $\varepsilon_\gamma \sim 1$, single-photon pair-produce with the magnetic field, resulting in low-energy secondary pairs, $\varepsilon_e \sim 1$.
The required magnetic field strength can be estimated by considering the energy gap between the adjacent Landau levels, $~\hbar \omega_B$, and comparing it to the electron rest mass energy, $\hbar \omega_B = m_e c^2$.
This results in $B_\mathrm{QED} = \frac{m_e^2 c^3}{e\hbar} \approx 4.4 \times 10^{13} \gauss$.
Before pair producing, the mean free path of photons in the magnetosphere is 
\footnote{Omitting a logarithmic function that weakly depends on the neutron star parameters.}
\begin{equation}
l_\gamma \sim \frac{R_c}{10} \frac{B_\mathrm{QED}}{B} \frac{1}{\varepsilon_\gamma}.
\end{equation}
This is $\sim R_c \sim R_\star$ for MeV photons.

As a result, the neutron star magnetosphere is quickly filled with $e^\pm$ that screen the initial voltage.
The creation of secondary pairs is halted when the longitudinal electric field, $E_\parallel$, is screened.
Therefore, considering a rotating magnetized ball in a vacuum is not enough since the magnetosphere is dynamic and capable of generating its own plasma.

This nonlinear feedback mechanism causes cyclic bursts of pair creation where a single seed electron can trigger a generation of a huge number of pairs.
The resulting charges are blown out relativistically into the surrounding space.
After the magnetosphere has cleared from the pairs, the voltage starts to grow again.
This leads to cyclic generation of bursts.

\subsection{Vacuum birefringence}

For extremely large magnetic fields, especially those of magnetars with $B\sim 10^{15}\gauss$, even the properties of the surrounding vacuum itself become modified.
This happens because of another quantum electrodynamic phenomenon called vacuum birefringence.

Birefringence, in general, is a property of, for example, crystals in which the speed of light inside the crystal is dependent on the direction of the light with respect to the axis of the crystal.
In crystals, this breaking of isotropicity can happen even if the crystal lattice is, for example, more tightly packed in one direction. 
This causes the crystal to have different refractive indices of light depending on the direction;
\textit{bi}refringence means that instead of one regular, isotropic refractive index, $n_1$, the medium has two indices, $n_1$ and $n_2$.

The vacuum around strongly magnetized neutron stars can become birefringent and develop a differing refractive index along and perpendicular to the local magnetic field direction, $n_\parallel$, and $n_\perp$.
The effect is caused by virtual electron-positron pairs of the vacuum background spontaneously appearing and annihilating.
Like any charged particle, these virtual pairs also gyrate in the strong magnetic field.
This creates an imaginary crystal lattice in the vacuum that modifies the propagation of electromagnetic waves through it.
Most notably, the polarization of the radiation is forced to align with the ambient magnetic field.

The strength of the effect can be quantified as \cite{2003MNRAS.342..134H},
\begin{equation}
n_\parallel - n_\perp = \frac{\alpha_{F}}{30\pi} \left( \frac{B}{B_\mathrm{QED}} \right)^2 \sin^2\alpha,
\end{equation}
where $\alpha_{F} \approx 1/137$ is the fine structure constants, and
 $\alpha$ is the angle between the direction of propagation and the external field.
The magnetic field remains strong enough to a distance known as the polarization limiting radius,
\begin{equation}
r_{\mathrm{pl}} \sim 10^{7} \left(\frac{B}{10^{12} \gauss}\right)^{2/5} \left(\frac{E_\gamma}{0.5 \kev}\right)^{1/5} (\sin\alpha)^{2/5} \cm.
\end{equation}
This renders the radiation in the inner regions, $r < r_{\mathrm{pl}}$, strongly birefringent and, therefore, also modifies the polarization of the radiation.
If the inner field is coherent, like a dipole field, the ordered magnetic field means that the polarization direction of the radiation is also coherent.
An observer integrating the polarized radiation over a large solid angle containing the face of the star will hence detect a large net polarization.
Therefore, a high degree of polarization is expected for radiation from magnetars.

\subsection{Superfluid and superconducting interiors}

The material at the core of the neutron star is thought to be in a superfluid state \cite{1969Natur.224..673B}.
The strong interaction force between the nucleons renders the charge-neutral neutrons into a superfluid state and the electrically charged protons into a superconducting state.
The electrons are expected to remain in their regular state because the corresponding superconducting transition temperature is very small compared to the temperatures relevant for neutron stars.

The superconductivity can lead to drastic changes in the magnetic properties of the object.
Most crucially, superconductors exhibit the Meissner effect, where magnetic flux is expelled from the superconducting regions.
However, the diffusion timescale of flux repulsion from macroscopic regions for the core material is very long ($t_D \sim 10^{22} \second$) owing to the enormous electrical conductivity (large density enables easy transport of charges).

The protons in the core are thought to form a type-II superconductor.
This state is characterized by a periodic array of quantized vortices of super-current, which is aligned parallel to the local magnetic field.
\footnote{
On the other hand, Type-I superconductors have fine-scale, alternative regions of normal material encapsulating magnetic flux and superconducting regions exhibiting the Meissner effect.
The state of the superconductor depends on the ratio of proton coherence length to the penetration depth.
}
The magnetic flux associated with an individual proton vortex is
\begin{equation}
    \phi_0 \approx \frac{h c}{2e} \approx 2\times10^{-7} \gauss \cm^2,
\end{equation}
and the number of vortices per area is, therefore, $B/\phi_0$, meaning that billions of super-current vortices can thread the core.
For $B \sim 10^{12}\gauss$, the vortex lattice has a lattice separation of roughly $5 \times 10^{-10}\cm$.
This is large compared to the inter-particle distance, meaning that the vortices are mesoscale in size.

The neutrons in the core are also in a superfluid state.
A salient feature of the superfluid state is that the angular momentum of the superfluid becomes quantized into vortices that carry the angular momentum.
Therefore, the core is expected to be composed of multiple vortices carrying the angular momentum of the rotating star.
These superfluid vortex lines are parallel to the rotation axis.
Each neutron pair vortex supports \ $\hbar$ of angular momentum; 
multiple vortices are, therefore, required to support the total angular momentum.

The existence of a superfluid component and the possibility of a superconducting core makes neutron stars the largest blobs of superfluid material in the Universe.
They allow us to study superfluidity and superconductivity at the extreme limit since the core can remain superfluid even up to temperatures of millions of Kelvin due to the immense pressure.

\subsection{Gliches and quakes}

The high accuracy of pulsar timing measurements makes it possible to measure even the tiniest change in the star's rotation rate.
These high-precision measurements have revealed a phenomenon called glitches. 
Glitches are a rare, short-duration pulsar timing phenomenon.
They are seen as sudden positive jumps of rotational frequency, $\Delta \nu >0 $ (determined as the difference between the frequency after and before the glitch), in the pulsar timing data, followed by a negative change of the slope, $\dot{\nu} < 0$.
The glitch magnitudes vary from $\Delta \nu/\nu \approx 10^{-12}$ to $10^{-5}$ and they have been detected in over $100$ rotation-powered pulsars and magnetars \cite{2011MNRAS.414.1679E}.

Most glitches are followed by a recovery phase, where the spin-down rate increases and tends toward the pre-glitch value.
The recovery phase can be modeled with an exponential function, $\propto \mathrm{e}^{-t/t_r}$, with a typical relaxation time of $\sim 100 ~\mathrm{days}$ together with a longer time-scale component with a relaxation time-scale of $\sim 1000~\mathrm{days}$.

Physically, the glitches are thought to occur because of a rapid transfer of angular momentum between the superfluid interior and the outer crust of the neutron star, driven by catastrophic unpinning of the superfluid vortices.
\cite{1975Natur.256...25A, 2015IJMPD..2430008H}
The angular momentum mismatch is created by slowing down the star's crust with electromagnetic torques.
The rotation of the superfluid interior of the star, on the other hand, remains unchanged because the quantized vortices remain pinned. 
After the rotation mismatch exceeds an unknown threshold value, the vortices are unpinned and begin to slide along the lattice plane.
Migration of the vortices away from the axis of rotation results in a spin-down of the superfluid component, bringing the two components closer to co-rotation.

Magnetars are observed to also anti-glitch, where sudden negative, $\Delta \nu < 0$, rotation rate changes, i.e., spin-downs (rather than conventional spin-ups) of the star are observed \cite{2013Natur.497..591A}.
These events are speculated to occur due to magnetospheric reconfiguration.

Something needs to trigger the catastrophic unpinning of the superfluid vortices.
The most plausible candidates are starquakes and catastrophic readjustments of the crust \cite{1976ApJ...203..213R}.
The quakes are expected to occur when the crust accumulates stress beyond a critical strain that it can sustain, leading to a mechanical failure of the crystal lattice \cite{2009PhRvL.102s1102H}.
Why or how such large stress can build up in the crust lattice is still unknown.

Seismic motions of the crust will also couple the pulsar interior and the magnetosphere.
The quake is expected to launch an elastic shear wave into the crust lattice, and since the magnetic field is frozen in the crust and liquid core, it also leads to perturbation of the magnetosphere.
The resulting crustal oscillations deform the magnetic field lines and excite Alfv\'en waves on the magnetosphere \cite[e.g.,][]{2020ApJ...897..173B}.
These waves can also interact with each other and heat the magnetospheric plasma \cite[e.g.,][]{2019ApJ...881...13L,2022PhRvL.128g5101N}.

\subsection{Giant bursts and fast radio bursts from magnetars}

A similar kind of star quake event, only more extreme, is thought to also power magnetar flaring phenomena.
In this case, super-strong crustal quakes are proposed as triggers of giant bursts \cite{1992ApJ...392L...9D}.
Similarly, smaller quakes could produce fast radio bursts as a side product of magnetospheric activity.

\textbf{Giant bursts.}
Giant bursts are massively energetic blasts of radiation.
A typical giant burst consists of a short, $\sim 0.2\second$ sharp pulse of hard X-rays and soft gamma-rays of energies $E \sim 100 \,\mathrm{keV}$ to $\mathrm{MeV}$.
The initial pulse is followed by a fainter, slowly-fading tail lasting minutes.
In the case of the famous March 5th, 1979 event, the tail also varied in intensity with a sinusoidal perturbation (with two differing peaks per cycle) and a period of $8\second$.
These oscillations were seen for more than 20 cycles.

The leading model that could power these flares is a large-scale magnetospheric reconfiguration --- these models rely on the large magnetic energy reservoir of the magnetar and on the fact that energy stored in the magnetic field can be quickly released to explain the extremely short duration of the pulses.
In this picture, a strong crustal quake launches Alfvén waves into the magnetosphere.
The waves twist the field lines, causing them to inflate and expand.
The inflated bundle over-twists and undergoes rapid magnetic reconnection. 
The magnetic reconnection leads to the creation of strong non-ideal electric fields, and turbulent plasma motions that can energize and heat the local plasma \cite[e.g.,][]{2021ApJ...921...87N}.
The mechanism resembles the energization mechanism observed behind solar flares.

The twisted, evolving bundle carries a strong electric current and a relativistic stream of particles along the arched magnetic line loop.
The streaming charged particles quickly radiate their energy as synchrotron radiation and can furthermore Compton up-scatter the photons up to relativistic energies.
These high-energy photons will pair-create, forming a trapped pair-plasma cloud.
Cooling of this pair plasma in the strong magnetic field provides a mechanism for generating the ultra-luminous soft gamma-ray burst seen from the soft gamma-ray repeaters.
Furthermore, the streaming plasma will also fall back to the star, heating the magnetic footprints.
Self-consistent modeling of the giant bursts requires detailed plasma physics models in strong magnetic fields, and large radiation densities \cite{2021ApJ...921...92B}.

\textbf{Fast radio bursts.}
Similar plasma-physics-related phenomena to giant bursts are the recently found fast radio bursts (FRBs) \cite{2019ARA&A..57..417C, 2019A&ARv..27....4P, 2022A&ARv..30....2P}.
FRBs are short $\sim 1\mathrm{ms}$ duration radio pulses with energies $\sim 10^{35}$ to $10^{43} \ergs$ (translating to $50 \,\mathrm{mJy}$ to $100 \,\mathrm{Jy}$) observed between radio frequencies from $400\,\mathrm{Mhz}$ to $8 \,\mathrm{GHz}$.
The emission is strongly polarized, exhibiting linear polarization degrees up to $100\%$.
FRBs are the biggest radio-emission-based enigma since the discovery of pulsars.
The latest coincident observation of FRB, together with a flaring magnetar, undeniably established a connection between the two (at least for some sub-class of FRBs).

The extremely high brightness temperature required to explain the FRB emission requires a coherent electromagnetic emission mechanism.
The high coherence means that many particles are required to emit in phase rendering the problem automatically into a plasma physics problem.
Current leading coherent emission mechanisms are the synchrotron maser instability from magnetized collisionless shocks \cite{2021PhRvL.127c5101S} and plasma waves (magnetosonic fast modes) from collapsing, reconnecting current sheets beyond the light cylinder of the magnetars \cite{2020ApJ...897....1L, 2022ApJ...932L..20M}.
These and other emission mechanisms are reviewed in \cite{2021Univ....7...56L}.

\subsection{Extreme particles: cosmic rays, neutrinos, and more}

Finally, it should be mentioned that neutron stars are thought to be the origin of many high-energy particle events like cosmic rays, neutrinos;
and, more speculatively, even sources of dark-matter candidates like axions.
Observations of all of these can be used to probe fundamental symmetries in physical laws because their energies are unreachable by Earthly laboratories.
In addition, they help us better understand the nature of particle acceleration in our Universe.

\textbf{Cosmic rays.}
Cosmic rays are relativistic protons or heavier atomic nuclei moving with velocities near the speed of light.
The flux of cosmic rays is constantly raining down on us on Earth.
They are typically detected when they slam into the upper layers of Earth's atmosphere and dissipate their energy in showers of secondary high-energy particles (like X-rays, protons, electrons, alpha particles, pions, muons, neutrinos, and neutrons). 
The secondary particles have been first observed decades ago and are now routinely monitored.
In the most extreme case, a dim fluorescent light from a cosmic ray exciting atmospheric nitrogen can be detected if the night is dark and moonless.
On a more regular basis, they are detected by space-borne observatories such Fermi Gamma-Ray Space Telescope (detecting the secondary gamma-rays from the showers) or ground-based observatories such as HESS or HAWCK (detecting Cerenkov radiation from the cosmic rays entering the Earth's atmosphere). 

The cosmic ray energy spectrum peaks at $\sim 10^9 \eV$ and extends into a broad power-law tail.
The power-law tail has two breaks:
first one, (named ``knee'') at $\sim 10^{15}\eV$ and another (named ``ankle'') at $\sim 10^{18}\eV$.
The lowest energy particles, $\lesssim 10^{10}\eV$, are thought to originate mostly from our own Sun (due to solar flares accelerating particles), while the intermediate energies, $10^{10} \eV \lesssim E_{\mathrm{CR}} \lesssim 10^{15} \eV$, are galactic cosmic rays, accelerated by shocks in supernova remnants \cite{2011ApJ...742L..30G, 2013Sci...339..807A}---neutron stars in action!
Other more distant sources of high-energy cosmic rays could include, for example, active galactic nuclei (supermassive black holes at the centers of galaxies).

The most energetic cosmic rays, named ultra-high-energy cosmic rays (UHECRs), can have kinetic energies up to $10^{20} \eV$.
In comparison, the LHC (Large Hadron Collider) on Earth can only probe energies up to $\sim 10^{17}\eV$.
These extreme particles are composed of the lightest elements, most likely pure protons or light nuclei like helium cores.
However, each carries a kinetic energy equivalent to a baseball ($m_b \sim 150\gram$) traveling with a velocity of $v_b \approx 100 \mathrm{km} \,\mathrm{h}^{-1} \sim 2800 \cm \second^{-1}$,
\begin{equation}
E_{\mathrm{kin}} = \frac{1}{2} m_b v_b^2 \sim \frac{1}{2} 150 \gram \times (2800 \cm \second^{-1})^2 \sim 6\times 10^{8} \erg \sim 4\times 10^{20} \eV.
\end{equation}
The UHECR particle is extremely relativistic since its energy is $E_{\mathrm{UCR}} = \gamma_{\mathrm{UCR}} m_{\mathrm{p}} c^2$, so that $\gamma_{\mathrm{UCR}} \sim 10^{10}$ for $E_{\mathrm{UCR}} \sim 10^9 \erg$.
Their energy is so high that they are theoretically expected to quickly lose it by scattering against the low-energy photons of the cosmic microwave background (Greisen-Zatsepin-Kuzmin limit; GZK limit).
This means that they cannot be produced by, for example, the most distant quasars or active galactic nuclei but must instead originate from somewhere nearby.
The big question is, what mechanism can accelerate protons to such extreme Lorentz factors in our neighborhood?
This mystery is complemented by the latest observation of PeV photons from the Crab \cite{2021Sci...373..425L} that also strongly challenges our picture of the particle acceleration and its limits.
Neutron stars are plausible engines for powering the particle acceleration in many cosmic ray generation models.


\textbf{Neutrinos.}
Another class of extreme particle is the neutrinos ---
light-weight ($m_\nu \lesssim 0.120 \eV$) subatomic particles that interact only via weak interaction and gravity.
Neutrinos come in three leptonic flavors: electron neutrinos ($\nu_\mathrm{e}$), muon neutrinos ($\nu_\mu$), and tau neutrinos ($\nu_\tau$).
They oscillate between the different flavors; a neutrino with a specific flavor is composed of a quantum superposition of all three flavors and can randomly change its appearance via flavor oscillations.
They are produced in radioactive decay, nuclear reactions, or as secondary particles by cosmic rays interacting with Earth's atmosphere.

Most of the neutrinos detected on Earth come from the Sun.
However, some neutrinos are confirmed to have a cosmic origin.
These ``cosmic neutrinos'' have been directly observed from a core-collapse supernova explosion SN 1987A 
\cite{1987PhRvL..58.1490H}, making supernovae and neutron stars prime candidates for neutrino sources.
The primary mechanism for neutrino emission in the supernova explosions (i.e., non-degenerate matter with density $\rho < 10^7 \gram \cm^{-3}$) is the annihilation of electron-positron pairs into neutrino-antineutrino pairs, $e^- + e^+ \rightarrow \nu + \bar{\nu}$ (where $\nu$ is the neutrino and $\bar{\nu}$ is anti-neutrino) \cite{1967ApJ...150..979B}.
At high temperatures, $T \gtrsim 10^9 \Kelvin$, the volumetric energy loss rate from the pair annihilation is (see Ref. \cite{2018MNRAS.476.2867M} for discussion),
\begin{equation}
    Q_{\nu} \approx 4\times10^{24} \left(\frac{T}{10^{10} \Kelvin}\right)^9 \erg \cm^{-3} \second^{-1}.
\end{equation}
Similar mechanisms could be at play in ultraluminous X-ray sources if they are interpreted as systems with super-Eddington accretion channeled onto magnetized neutron stars \cite{1976MNRAS.175..395B, 2014Natur.514..202B};
the high mass flux, production of an accretion shock, and the resulting shock-heated high temperatures would render the accretion columns on top of these neutron stars into efficient sources of neutrinos --- neutrino pulsars \cite{2018MNRAS.476.2867M}!

Detection of astrophysical neutrinos holds great promise for probing fundamental physics.
In the event of a close-by galactic supernova, there is hope that we could accurately measure the neutrino energies and flavor spectrum.
This would help us better constrain the core-collapse supernova physics and any high-energy non-standard interactions beyond the Standard Model of particle physics.

\textbf{Axions.}
Neutron stars are also speculated to work as probes of various dark matter candidate particles \cite{2021Univ....7..401T}.
The most intriguing of these is the axion, an extremely light ($m_a < 1 \eV$) elementary particle, postulated to resolve the strong CP symmetry problem in quantum chromodynamics.
Axions couple to the electromagnetic fields and can change the behavior of the electrodynamic processes in the magnetosphere.
Most notable of these couplings is the alteration of the physics at the magnetospheric gap region, $\vec{E}\cdot\vec{B} \ne 0$, which is associated with a strong voltage, efficient production of gamma-rays, and therefore also the possible origin of axions \cite{2021PhRvD.104e5038P}.

\section{Summary}

Research of neutron stars encompasses many branches of modern physics, from gravitational physics to nuclear and particle physics, from electrodynamics to plasma physics.
What makes neutron stars unique is that they exhibit extreme behavior in each one of these fields, often offering novel insight into the limits of our physical theories. 

We started by presenting a (short) history of neutron stars and their discovery. 
For a more thorough historical discussion on neutron stars, we refer the reader to \cite{yakovlev_2013}.
We then reviewed the basic arguments behind the studies of gravitation, nuclear \& particle physics, electromagnetism and \& quantum electrodynamics, and plasma physics in the context of neutron stars.
Each of these fields is still full of mysteries and unexplained phenomena that need to be solved to understand the physics and astrophysics of neutron stars better.

In this chapter we have provided a quick precursor to the field of fundamental physics of neutron stars, and for the interested reader more thorough reviews of these topics can be found from the following studies.
Excellent sources of more information for studies of gravity include \citep{misner_1973, paschilidis_2017}.
The physics of emergent radiation is discussed in \citep{potekhin_2014} and the effects of light-bending in \citep{nattila_2018, suleimanov_2020}.
Notably, a simple model of light bending in curved spacetime is developed in \citep{beloborodov_2002, poutanen_2020}.
Studies of the neutron star equation of states are reviewed in \citep{ozel_2016rev, lattimer_2019}.
Measurements of neutron star radii are discussed in \cite{miller_2016, suleimanov_2016, degenaar_2018}.
The recent gravitational wave observations are presented, for example, in \citep{baiotti_2019}.
Studies of neutron stars and pulsar magnetospheres are reviewed in \citep{cerutti_2017, beskin_2018}.
Pair cascades and quantum electrodynamic effects are discussed in \citep{svensson_1982a, svensson_1982b}.
Magnetars and their observations are reviewed in \cite{2017ARA&A..55..261K}, and
plasma physics of fast radio bursts are well-discussed in \cite{2021Univ....7...56L}.

\section{Cross-References}

\begin{itemize}
    \item Low mass X-ray binaries (Bahramian \& Degenaar)
    \item Fundamental physics with black holes (Carcia)
    \item Isolated neutron stars (Borghese)
    \item Low magnetic field (Di Salvo, Burderi, Iaria)
    \item High magnetic field (Mushtukov)
\end{itemize}

\end{document}